\documentclass[a4paper, 11pt]{article}
\usepackage{amsfonts, amsmath, amssymb, amsthm, amscd, setspace, subfigure,mathtools,stmaryrd}
\usepackage{verbatim, geometry,mdframed,graphicx,enumitem,cancel}
\usepackage{pictexwd,dcpic,float,array,slashed,fancyhdr}
\usepackage[utf8]{inputenc}
\usepackage[dvipsnames]{xcolor}
\usepackage[all,cmtip]{xy}
\usepackage{xr-hyper}
\usepackage{hyperref}
\usepackage[doi=false,eprint=false,maxbibnames=4,maxcitenames=4]{biblatex}
\DeclareBibliographyDriver{unpublished}{
\usebibmacro{bibindex}
\usebibmacro{begentry}
\usebibmacro{author}
\setunit{\labelnamepunct}\newblock
\usebibmacro{title}
\newunit
\printlist{language}
\newunit\newblock
\usebibmacro{byauthor}
\newunit\newblock
\printfield{howpublished}
\newunit\newblock
\printfield{note}
\newunit\newblock
\usebibmacro{location+date}
\newunit\newblock
\usebibmacro{doi+eprint+url}
\newunit\newblock
\usebibmacro{addendum+pubstate}
\setunit{\bibpagerefpunct}\newblock
\usebibmacro{pageref}
\usebibmacro{finentry}}

\newcommand{\T}{\mathbf{T} _\mathcal{A}   }
\newcommand{\dbb}{ \mathbf{B}  _\mathcal{A} }
\newcommand{\dbbb}{ \mathbf{P}  _\mathcal{A} }
\newcommand{\p}{ \tilde p}
\newcommand{\ssigma}{\boldsymbol{\sigma}}

\newcommand{\FF}{F _{ \mathrm{SD} }}
\newcommand{\FFasd}{F _{ \mathrm{ASD} }}
\newcommand{\hsp}{ \varphi _+  }
\newcommand{\hsm}{ \varphi _{- } }
\newcommand{\hspm}{ \varphi _{\pm } }

\newcommand{\V}{\upsilon }

\newlength{\mymathln}

\begin{document}

\baselineskip 18pt
\begin{center}
{\Large \bf Harmonic Forms and Spinors on  the Taub-bolt Space}

\vspace{0.6cm} 
{\bf Guido Franchetti } \\
Dipartimento di Matematica Giuseppe Peano \\
Universit\`a di Torino\\  10123 Torino, Italy. \\
{\tt guido.franchetti@unito.it}

\end{center}
\baselineskip 14pt

\vspace{0.1cm} 

\begin{abstract}
\noindent 
This paper studies the space of $L ^2 $ harmonic forms and $L ^2 $ harmonic spinors on Taub-bolt, a Ricci-flat Riemannian 4-manifold of ALF type. We prove that the space of harmonic square-integrable 2-forms on Taub-bolt is 2-dimensional and construct a basis. We explicitly find all $L ^2 $  zero modes of $\slashed {D} _\mathcal{A} $, the Dirac operator twisted by an arbitrary $L ^2 $ harmonic connection $\mathcal{A}$, and    independently compute  the index of $\slashed {D} _\mathcal{A} $. We compare our results with those known in the case of  Taub-NUT and Euclidean Schwarzschild as these manifolds present interesting similarities with Taub-bolt. In doing so, we  generalise known results concerning harmonic spinors on Euclidean Schwarzschild. 
\end{abstract}
\newpage 
\tableofcontents
\newpage

\section{Introduction}
The aim of this paper is to study  $L ^2 $ harmonic forms and $L ^2 $  harmonic spinors on the Taub-bolt (TB) manifold, and to compare these structures with the corresponding ones on self-dual Taub-NUT (TN) and Euclidean Schwarzschild (ES).

The harmonic cohomology of non-compact manifolds is an interesting  topic since the usual Hodge decomposition results do not apply. In fact, little is known about generic harmonic forms on a non-compact manifold, but the situation improves in the case of square-integrable ($ L ^2 $) ones. In particular, we will make use of results \cite{Hausel:2004vv} relating the harmonic cohomology of a Riemannian manifold $M$ satisfying certain asymptotic conditions with the ordinary cohomology of a particular compactification of $M$, which we refer to as the HHM compactification.

From the physical viewpoint, most interesting models of spacetime are non-compact and harmonic forms on them represent solutions of the Maxwell equations of electrodynamics. The $L ^2 $ assumption is a natural finite-energy condition. TB, TN and ES are all Euclidean continuation of general relativity solutions, hence the problem we are studying is the Euclidean continuation of one having physical relevance.

On a Riemannian 4-manifold we can ask for the electromagnetic field strength, a 2-form, to be self-dual. Since electromagnetism is an Abelian theory, self-dual closed forms are sometimes referred to as Abelian instantons. An interesting property of Abelian instantons is that they give a vanishing contribution to the stress energy tensor. We will prove that the space of $L ^2 $ harmonic 2-forms on TB and ES is 2-dimensional,  on TN is 1-dimensional, and in all cases self-dual 2-forms comprise a 1-dimensional subspace. 

Of course, some of these results are well known. In particular, harmonic forms on ES have been considered in \cite{Etesi:432418,Pope:1978ffa,Jante:2016vq}. While \cite{Pope:1978ffa,Jante:2016vq} focus their attention on the self-dual 2-forms, in \cite{Etesi:432418} the 2-dimensional space of harmonic 2-forms is explicitly constructed making use of the fact that any harmonic form on ES has to be rotationally invariant. We construct the space of $L ^2 $ harmonic forms on ES following a different approach as we find the relation between harmonic cohomology and HHM compactification to be a particularly interesting one. The space of harmonic $L ^2  $ forms on TN is generated by a self-dual one which has been studied before \cite{Pope:1978ffa,Hitchin:398393,Jante:2014ho,Franchetti:2014ue}. The case of TB has not been considered before. 

Harmonic spinors, that is solutions of the (massless) Dirac equation, on a Riemannian manifold are another topic of both mathematical and physical interest. A simple argument based on Lichnerowicz's formula and the fact that TN, ES have vanishing scalar curvature shows that they admit no non-trivial $L ^2 $ harmonic spinors. As for TB, it is not a spin manifold. It is therefore natural to look at the twisted Dirac equation, obtained by coupling the Dirac operator to an Abelian connection. Physically, we are coupling the fermion described by the spinor field to an electromagnetic field.

In order for the problem to retain its interest, the Abelian connection should not be completely arbitrary and connections having $L ^2 $ harmonic curvature are  natural candidates. Harmonic spinors on TN have been studied in \cite{Pope:1978ffa,Jante:2014ho}; harmonic spinors on ES in \cite{Pope:1978ffa,Jante:2016vq}. We generalise the known results on ES by allowing for an arbitrary, rather than self-dual, $L ^2 $ harmonic connection. The case of TB has not been considered before.

As discussed, this paper contains no new results as far as TN is concerned, and only moderately new ones in the case of ES. In light of this fact, results  concerning   TN  are recalled without being re-derived, and harmonic spinors on  ES  are treated succinctly making heavy use of the treatment in \cite{Jante:2016vq}. By doing so, we  focus our presentation on the novel case of TB. Even so, one may  wonder why considering TN and ES at all. We would like to make the case that a comparison of this three manifolds in a single place is worth having.

TB, TN and ES  are all  rotationally symmetric metrics (that is, their isometry group has an $SO (3) $ or $SU (2) $ subgroup) which admit an additional  isometric $U (1) $ action. In fact, they all are special cases of a 1-parameter family of metrics, the non self-dual Euclidean Taub-NUT. The $U (1) $-action has fixed points: a single one, called a nut, in the case of TN, a 2-sphere, known as a bolt, in the case of TB and ES. The  $SU (2) $ action on TB and TN has orbits  homeomorphic to  a Hopf fibration, hence a twisted $U (1) $ bundle over $S ^2 $. On ES this twisted bundle  is replaced by the trivial fibration $S ^2 \times S ^1 $.

As we can see, TB  displays the non-trivial topology of ES (which retracts onto the bolt), and the non-trivial $U (1) $ fibration of TN. Because of these similarities and differences, a comparison with TN and ES  is in order. As expected, $L ^2 $ harmonic forms and spinors on TB reflect its more intricate structure.

Besides TB, TN and ES, the non self-dual Euclidean Taub-NUT  family  contains a 4th smooth metric, the  Eguchi-Hanson (EH) one.  The reason for focusing on TN, TB and ES is that they all are of ALF type, that is their asymptotic volume growth  resembles that of Euclidean 3-space, while EH is of ALE type, that is its asymptotic growth resembles that of Euclidean 4-space. TN and EH are hyperk\"ahler manifolds hence in particular Ricci-flat and self-dual. TB and ES are also Ricci-flat but not self-dual.

The structure of this paper is as follows. In Section \ref{metrics} we introduce TB, TN and ES as particular cases of the non-self dual Euclidean Taub-NUT space. In Section \ref{harforms} we fully describe the space of $L ^2 $ harmonic forms on them. In particular, we identify their HHM compactification up to homeomorphisms, and relate  harmonic 2-forms to the Poincar\'e duals of non-trivial 2-cycles of this compactification. Section \ref{hspinors} is devoted to zero modes of the Dirac operator twisted by a connection having harmonic curvature. We explicitly find a family of solutions. By comparing their number with the calculation of the index of the twisted Dirac operator in Section \ref{apsthmasf} we can conclude that, if the twisting connection has self-dual curvature, we have found all the $L ^2 $ harmonic spinors. In the case of a generic harmonic connection, the number of solutions we find is still equal to the index, but in the absence of a vanishing theorem for either the kernel or cokernel we cannot exclude the existence of other solutions. The results obtained in Sections  \ref{harforms} and \ref{hspinors} are discussed and compared in Section \ref{conclusions}.

\newpage 
\section{The non-self-dual Euclidean Taub-NUT manifold}
\label{metrics} 
The analytic continuation to Euclidean signature of the Lorentzian Taub-NUT manifold has a metric of bi-axial Bianchi IX type
\begin{equation}
\label{b9} 
g = f ^2 \mathrm{d} r  ^2 + a ^2 (\eta _1 ^2 + \eta _2 ^2 ) + c ^2 \eta _3 ^2,
\end{equation} 
with $\eta _i $ left-invariant 1-forms on $SU (2) $,
\begin{equation}
\label{livforms} 
\begin{split}
\eta _1 &= \sin \psi  \, \mathrm{d} \theta - \cos \psi \sin \theta \, \mathrm{d} \phi,\\
\eta _2 &= \cos \psi \, \mathrm{d} \theta + \sin \psi \sin \theta \, \mathrm{d} \phi ,\\
\eta _3 &= \mathrm{d} \psi + \cos \theta \, \mathrm{d} \phi,
\end{split}
\end{equation} 
\begin{equation}
\begin{split} 
f &=-\sqrt{ \frac{(r  - N )(r + N )}{( r - r _+ ) ( r - r _-  )}}, \quad r _{ \pm } = M \pm \sqrt{ M ^2 - N ^2 },\quad 
 a = \sqrt{ r ^2 - N  ^2}, \quad c =-\frac{2N}{f}.
\end{split} 
\end{equation} 
It is a Ricci-flat metric with  isometry group $SU (2) \times U (1) $ which has been considered in detail in \cite{Page:1978hc}. 

Up to a change of orientation,  $M \geq N \geq 0 $, so that $r _+ \geq r _- $.
In order to avoid curvature singularities, $r \in [r _+ , \infty )$ and for $r > r _+ $ fixed $r$ hypersurfaces have the topology of a lens space with $s$ points identified. For generic values of $M$ and $N$, the surface $r =r _+ $ is a 2-sphere  of fixed points of the $U (1) $ isometry generated by the Killing vector field $\partial / \partial \psi $, known as a bolt. The angles $( \theta , \phi ) \in [0, \pi ]\times [0, 2 \pi )$ are the usual coordinates on a 2-sphere.
In order to avoid conical singularities along the bolt, the angle $\psi$ has to be in the range $ \psi \in [0, 4 \pi /s )$ with $s$ a positive integer.
 It follows from the analysis in \cite{Page:1978hc} that only a few values of the parameters $M $, $N$, $s$, to be reviewed below, yield smooth manifolds.

We take the orthonormal coframe
\begin{equation}
\label{coframe} 
e ^1 =a \eta _1 , \quad e ^2 =a \eta _2 , \quad e ^3 =c \eta _3 , \quad e ^4 = -f \, \mathrm{d} r 
\end{equation} 
and  the volume element
\begin{equation}
\begin{split} 
\operatorname{vol} &=e ^1 \wedge e ^2 \wedge e ^3 \wedge e ^4 
= - 2N ( r ^2 - N ^2 ) \sin \theta \, \mathrm{d} \psi \wedge  \mathrm{d} \theta \wedge \mathrm{d} \phi \wedge \mathrm{d} r .
\end{split} 
\end{equation} 

\subsection{Taub-bolt}
The case $M =\tfrac{5}{4} N $, $s =1 $ gives the Taub-bolt (TB) manifold, which is topologically equivalent to $\mathbb{C}  P ^2 $ with a point removed, or equivalently to a disk bundle over a 2-sphere, the bolt $r =r _+ $. Its de Rham cohomology is therefore
\begin{equation} 
H ^p _{ \mathrm{dR} }(\mathrm{TB}  ) =
 \begin{cases} \mathbb{R}  &\text{if $ p =0,2 $},\\ 0 &\text{otherwise}.   \end{cases}
\end{equation}
The topology of a fixed $r$ hypersurface is a Hopf fibration. 

The metric is 
\begin{equation}
\label{tbm} 
g _{  \mathrm{TB} }  =\left( \frac{r ^2 - N ^2  }{(r -2 N)(r-N/2) } \right) \mathrm{d} r ^2 + (r^2 - N ^2 ) \left( \eta _1 ^2 + \eta _2 ^2 \right) 
+4 N ^2 \left( \frac{(r -2 N)(r-N/2)}{r ^2 - N ^2 } \right) \eta _3 ^2 ,
\end{equation} 
with $r \in [ 2 N, \infty ) $, $\psi \in [0, 4 \pi )$ and
\begin{equation}
\label{tbvalues} 
a _{ \mathrm{TB} }= \sqrt{ r ^2 - N ^2 }, \quad c _{ \mathrm{TB} }=2 N \sqrt{ \frac{(r - 2N)(r-N/2) }{r ^2 - N ^2  }}, \quad f _{ \mathrm{TB} } = - \frac{2N}{c_{ \mathrm{TB} }}.
\end{equation}

\subsection{Taub-NUT}
The case $M =N $, $s =1 $ gives the self-dual  Taub-NUT manifold, which from now on we refer to simply as Taub-NUT (TN). In this special case, $ r _+ =r _- = N $ and $r = N $ is a single point, known as a nut. The resulting manifold is diffeomorphic to $\mathbb{R}  ^4 $. The topology of a fixed $r$ hypersurface is a Hopf fibration as in TB. 

The metric is 
\begin{equation}
\label{tnm} 
g _{  \mathrm{TN} }  =\left( \frac{r + N }{r - N } \right) \mathrm{d} r ^2 + ( r ^2 - N ^2 ) \left( \eta _1 ^2 + \eta _2 ^2 \right) +4 N ^2\left(  \frac{r - N }{r + N }\right) \eta _3 ^2 .
\end{equation} 
so that
\begin{equation}
a _{ \mathrm{TN} }= \sqrt{ r ^2 - N ^2 }, \quad c _{ \mathrm{TN} }=2 N \sqrt{ \frac{r - N }{r + N }}, \quad f _{ \mathrm{TN} } = - \frac{2N}{c_{ \mathrm{TN} }},
\end{equation} 
with $r \in [N, \infty )$, $\psi \in [0, 4 \pi )$.

\subsection{Euclidean-Schwarzschild}
Two limiting cases of the non-self-dual Taub-NUT metric also yield smooth manifolds. For $s =2 $, $N \rightarrow \infty $ with  $r _+ ^2 - N ^2 $ converging to a positive constant, one obtains the Eguchi-Hanson  space which we will not consider here.

Making the coordinate change $\psi ^\prime = 2 N \psi \in [ 0, 8 \pi N /s)$ and taking the limit $ N \rightarrow 0 $, $ s \rightarrow 0 $, $N / s \rightarrow M $ one obtains the Euclidean-Schwarzschild (ES) manifold, which is topologically  a disk bundle over a 2-sphere. Therefore, its de Rham cohomology is, as for TB, that of a 2-sphere. To the contrary of TN and TB however, the topology of a fixed $r$ hypersurface is a trivial circle bundle over $S ^2 $.

The metric is
\begin{equation}
g _{ \mathrm{ES} } 
= \frac{ \mathrm{d} r  ^2 }{1 - \tfrac{2 M }{r }}  + 16 M ^2\left( 1 - \tfrac{2 M }{r } \right)   \mathrm{d} \chi   ^2 + r  ^2 \mathrm{d} ( \eta _1 ^2 + \eta _2 ^2 ),
\end{equation} 
with $\chi =\psi ^\prime /(4 M ) \in [0, 2 \pi )$, $r \in [2 M , \infty )$. To uniform the notation we write 
\begin{equation}
g _{ \mathrm{ES} }  = f _{ \mathrm{ES} }^2 \mathrm{d} r ^2  + a_{ \mathrm{ES} } ^2 (\mathrm{d} \eta _1 ^2 + \mathrm{d} \eta _2 ^2 )+   c _{ \mathrm{ES} }^2 \, \mathrm{d} \chi ^2,
\end{equation} 
and $e ^1 = a_{ \mathrm{ES} } \, \eta _1 $,  $e ^2 = a_{ \mathrm{ES} }\,  \eta _2 $, 
$e ^3 = c_{ \mathrm{ES} }\,  \mathrm{d} \chi  $, $e ^4 = - f_{ \mathrm{ES} }\,  \mathrm{d} r$, with
\begin{equation}
a _{ \mathrm{ES} }= r, \quad c _{ \mathrm{ES} }=4M \sqrt{ 1 - \frac{2M }{r  }}, \quad f _{ \mathrm{ES} } = - \frac{4M}{c_{ \mathrm{ES} }}.
\end{equation}

\newpage 
\section{Harmonic cohomology}
\label{harforms} 
In \cite{Hausel:2004vv}, the space $L ^2 \mathcal{H} ^p (M) $  of harmonic $L ^2 $ differential $p$-forms on a non-compact manifold $M$ satisfying certain asymptotic conditions is related to the ordinary cohomology of a particular compactification $X _M $ of $M$, which we refer to as the Hausel-Hunsicker-Mazzeo (HHM) compactification of $M$. 

More precisely, let $(\overline{ M },g) $ be a smooth compact Riemannian manifold with boundary. We assume that $\partial \overline{ M } $ is the total space of a fibration $ F \hookrightarrow \partial \overline{ M } \xrightarrow{ \pi } B $. The space $\overline{  M} $  is said to have a fibred boundary metric if there is a neighbourhood $U$ of $\partial \overline{  M} $ where 
\begin{equation}
g = \mathrm{d} r ^2 + r ^2 \tilde h + k,
\end{equation} 
with  $x =1/r $ a boundary defining function, i.e.~$ x =0 $, $\mathrm{d} x \neq 0 $ on $\partial \overline{  M} $, $ \tilde h $  a smooth extension to $U$ of $ \pi ^\ast h $ for $h$ some metric on $B$, and $k$ a symmetric 2-tensor which restricts to a metric on $F$.  

Let  $M$  be the non-self-dual Taub-NUT space with metric (\ref{b9}),  $\overline{ M } _R $ a truncation of $M$ at finite radius $ R > r _+ $. 
The hypersurface $\partial \overline{ M } _R$ of fixed $r =R $  is a  fibration $ S ^1 \hookrightarrow \partial \overline{ M } _R \rightarrow \Sigma _R  $. For ES $\partial \overline{ M } _R   $ is the trivial fibration $S ^2 \times S ^1 $ while for TN and TB it is the Hopf fibration. In all cases $ \Sigma _ R  \simeq S ^2 $.
The HHM compactification is obtained by collapsing the fibres of $\partial \overline{ M } _R$ in the limit $R \rightarrow \infty $. Note that $M =X _M \setminus \Sigma _\infty $.

For large $R$,  in a neighbourhood of $\partial \overline{ M } _R  $ the metric (\ref{b9}) is, to leading order,
\begin{equation}
g \sim \mathrm{d} r ^2 + r ^2 ( \eta _1 ^2 + \eta _2 ^2 ) + 4 N ^2 \eta _3 ^2 ,
\end{equation} 
hence of fibred boundary type.  
It is known \cite{Hausel:2004vv} Corollary 1, \cite{Franchetti:2014ue}, that in the case of a fibred boundary metric with 3d boundary fibred by spheres $X _M $ is a smooth compact manifold and its harmonic cohomology is given by
\begin{equation}
L  ^2 \mathcal{H} ^p (M ) =\begin{cases} 
H ^1_{ \mathrm{dR}}  (X _M , \Sigma _\infty ) &\text{if $p =1 $},\\  H ^2 _{ \mathrm{dR} } ( X _M ) &\text{if $p =2 $} ,\\ 0 &\text{otherwise} . \end{cases} 
\end{equation}

Let us determine $ H ^p _{\mathrm{dR}}(X _M )$ for  $M$ either TN, TB or ES. We take open sets $ U = X _M \setminus \Sigma _\infty   \simeq M $, $V$ a neighbourhood of $\Sigma _\infty  $. The open set $V$  is homotopically equivalent to $\Sigma _\infty =S ^2  $. The intersection $U \cap V $ retracts onto a hypersurface of large $r$ hence   $U \cap V \sim S ^3 $ for TN and  TB,  $U \cap V \sim S ^2 \times S ^1 $ for ES. The de Rham cohomology of TN (TB and ES) is that of a point (a 2-sphere).  A Mayer-Vietoris sequence applied to $U$, $V$, $U \cap V $ then gives
\begin{equation}
H ^p  _{ \mathrm{dR} }(X_\mathrm{TN} )= \begin{cases} \mathbb{R}  & \text{for $p =0,2,4 $},\\ 0 &\text{otherwise}, \end{cases}  \quad 
H ^p  _{ \mathrm{dR} }(X_\mathrm{TB} )=H ^p  _{ \mathrm{dR} }(X_\mathrm{ES} )= \begin{cases} \mathbb{R}  & \text{for $p =0,4 $},\\ \mathbb{R}  ^2 &\text{for $p =2 $},\\  0 &\text{otherwise.} \end{cases} 
\end{equation} 
It follows that harmonic cohomology is non-trivial only in degree 2, where
\begin{equation}
L ^2 \mathcal{H} ^2 (\mathrm{TN }) \simeq \mathbb{R},\qquad  
L ^2 \mathcal{H} ^2 (\mathrm{TB }) \simeq L ^2 \mathcal{H} ^2 (\mathrm{ES }) \simeq \mathbb{R} ^2 .
\end{equation} 

While the de Rham cohomology of $X _M $ is enough to determine the dimension of $L ^2 \mathcal{H} ^2 (M)   $, we can actually identify the HHM compactifications of TN, TB and ES up to homeomorphism.
In order to do so, we look at the  intersection forms of $X _{ \mathrm{TN}}$, $X _{ \mathrm{ TB}} $ and $ X _{ \mathrm{ES}} $. While on TB $\Sigma _\infty $ and the bolt, which we denote by $\Sigma _ \mathrm{b} $, generate the middle dimension homology, on ES, due to the triviality of the fibration, the bolt  and $\Sigma  _\infty $ belong to the same homology class.

In order to obtain a basis of $H _2 ( X _ \mathrm{ES}, \mathbb{A}  )$  on ES, we take $\Sigma _\infty $ and the surface $\Sigma _  \mathrm{c} $ constructed as follows.  Consider a line in ES connecting a point on the bolt with one on $\Sigma  _\infty $ along a line of constant $\theta $, $ \phi $.  This surface is parametrised by the $(r, \chi )$  coordinates and we take the orientation $ \mathrm{d} \chi \wedge \mathrm{d} r $.  Above each point except the endpoints lies a circle, hence the surface so constructed is topologically a 2-sphere. 

With respect to the basis  $ \{\Sigma _ \mathrm{b} , \Sigma _\infty \} $ on $ X _{ \mathrm{ TB}}$, $ \{\Sigma _\infty \} $ on $X _{ \mathrm{ TN}} $, $ \{ \Sigma_{ \mathrm{c}} , \Sigma _\infty \}$  on $X _{ \mathrm{ ES }} $, the intersection forms are
\begin{equation}
\label{intforms} 
Q _{ X_\mathrm{TB}  } = \begin{pmatrix}
- 1 &0 \\
0 & 1
\end{pmatrix} , \quad
Q _{X_ \mathrm{TN}  }=(1), \quad 
Q _{ X_\mathrm{ES}  } = \begin{pmatrix}
0 &1 \\
1 & 0
\end{pmatrix} .
\end{equation}

A quick application of van Kampen's theorem to the open sets $U $, $V$, $U \cap V $ considered above shows that in all cases $X _M $ is simply connected. By Freedman's classification results, see e.g.~\cite{Scorpan:2005wg}, the homeomorphism type of a \emph{smooth} simply connected 4-manifold is completely  determined by its intersection form. It follows that, up to homeomorphisms,
\begin{equation}
X _{ \mathrm{TN } } = \mathbb{C}  P ^2 , \quad 
X _{ \mathrm{TB } } = \mathbb{C}  P ^2 \# \overline{ \mathbb{C}  P ^2 }, \quad 
X _{ \mathrm{ES } } = \mathbb{C}  P ^1 \times \mathbb{C}  P ^1  ,
\end{equation} 
where $\overline{ \mathbb{C}  P ^2 }$ denotes $\mathbb{C}  P ^2 $ with the reversed orientation and  intersection form $(-1) $.

We  now determine a basis for $L ^2 \mathcal{H} ^2$ comprising of the Poincar\'e duals of the 2-cycles generating the homology of $ X _ \mathrm{TB}$, $X _ \mathrm{ TN} $, $X _ \mathrm{ ES} $.
 We recall that, see e.g.~\cite{Bott:1982tn}, if $M$ is a smooth $n$-manifold and $S$ a closed oriented $k$-submanifold, the \emph{(closed) Poincar\'e dual} of $S$ is the unique cohomology class $[\delta _S ]\in H ^{ n-k }_{ \mathrm{dR}}  (M) $ such that for any form $ \omega \in H _{ \text{cpt}} ^k (M) $, the cohomology group of $k$-forms on $M$ with compact support, we have
\begin{equation}
\int _S \omega = \int _M \omega \wedge \delta _S .
\end{equation}

We also recall that  if $\xi $ is a Killing vector field on a Ricci-flat manifold, then $ \mathrm{d}  \xi ^\flat $ is coclosed, hence  harmonic.\footnote{In the compact setting $\mathrm{d} \xi ^\flat $ would be trivial, but  we are interested in non-compact manifolds.} The 2-form   $* \mathrm{d} \xi ^\flat $ is also harmonic since the Hodge Laplacian commutes with $* $. We take 
\begin{equation}
\xi = \frac{ 1}{4 N^2  }\,  \frac{ \partial}{ \partial \psi },
\end{equation}
 the generator of the $U (1) $ isometry given by translation along the circle fibres. We have
\begin{equation}
\xi ^\flat = \frac{ c ^2}{4 N ^2 } \eta _3  =\left(  \frac{r ^2 - 2 M r + N ^2 }{r ^2 - N ^2 } \right) \eta _3 ,
\end{equation}
\begin{equation}
\begin{split} 
\label{dxigen} 
\mathrm{d} \xi ^\flat &= \frac{1}{4 N ^2  } \Big(2 c c ^\prime \mathrm{d} r \wedge \eta _3 - c ^2 \eta _1 \wedge \eta _2 \Big )
=\frac{1}{4 N ^2 } \Big(\frac{2 c ^\prime }{f} e ^3 \wedge e ^4 - \frac{c ^2 }{a ^2 } e ^1 \wedge e ^2 \Big),\\
* \mathrm{d} \xi ^\flat &=\frac{1}{4 N ^2 } \Big( \frac{ 2  c ^\prime a ^2 }{f}\eta _1 \wedge \eta _2  +  \frac{ c ^3 f}{a ^2 }  \eta _3 \wedge \mathrm{d} r\Big)
=\frac{1}{4 N ^2 } \Big(\frac{2 c ^\prime }{f} e ^1 \wedge e ^2 - \frac{c ^2 }{a ^2 } e ^3 \wedge e ^4\Big).
\end{split} \end{equation}

From (\ref{dxigen}) we see that
 $\mathrm{d} \xi ^\flat $ is  self-dual provided that
\begin{equation}
\label{sddpsi} 
\frac{c ^2 }{a ^2 } =- \frac{2 c ^\prime }{f}.
\end{equation} 
This condition should be compared with the equations for the self-duality of the Riemann tensor on a Bianchi IX metric
$
g =f ^2 \mathrm{d} r ^2 + a ^2 \eta _1 ^2 + b ^2  \eta  _2 ^2 + c ^2 \eta _3 ^2 ,
$
which are \cite{Gibbons:1979tb}
\begin{equation}
 \frac{ 2 bc }{f} a ^\prime = (b- c) ^2 - a ^2  +  2 \kappa\,   bc
\end{equation} 
and the  other two cyclic permutations of $ (a,b,c) $. Here $\kappa$ is a constant equal to 0 or 1.
For a bi-axial Bianchi IX metric $a =b $ and two equations reduce to
$\tfrac{2 a ^\prime }{f}=\tfrac{c}{a}  + 2( \kappa -1 )$,
while the remaining one becomes
\begin{equation}
\label{sdsixth} 
 \frac{ 2  c ^\prime}{f} =2 \kappa  -\frac{  c ^2 } {a ^2 }
\end{equation} 
which for $\kappa =0 $ is equal to (\ref{sddpsi}).

\subsection{Taub-bolt}
\label{hftb} 
In the case of TB
\begin{equation}
\mathrm{d} \xi ^\flat =  \mathrm{d} \left(\frac{(r- 2 N ) (r - N/2) }{r ^2 - N ^2 }  \, \eta _3 \right) 
=- \frac{ U ^\prime}{2N} e ^3 \wedge e ^4  - \frac{U}{r ^2 - N ^2 } e ^1 \wedge e ^2 
\end{equation} 
with
\begin{equation}
\label{ufunct} 
U = \frac{1}{f ^2 } =\frac{(r - 2 N ) (r - N /2 )}{r ^2 - N ^2}, \quad U ^\prime =\frac{N }{2} \left( \frac{5 N ^2-8 N r +5 r ^2 }{(r ^2 - N ^2 )^2 }\right) .
\end{equation} 
The form $\mathrm{d} \xi ^\flat $ is exact. It is not self-dual and  $ * \mathrm{d} \xi ^\flat $  is also harmonic but not exact. 

The bolt $ \Sigma _\mathrm{b}  $ and $\Sigma _\infty $ provide a basis for  $H _2 (X _{ \mathrm{TB} }, \mathbb{Z}  )$. It can be checked that, for
\begin{equation}
\label{fffftb} 
F ^{ \mathrm{TB}}_\infty = - \frac{\mathrm{d} \xi ^\flat}{4 \pi }  , \qquad 
F ^{ \mathrm{TB}}_{ \mathrm{bolt} } = \frac{5}{3} F ^{ \mathrm{TB}}_\infty - \frac{4}{3} * F ^{ \mathrm{TB}}_\infty,
\end{equation} 
we have
\begin{equation}
\label{tbvarint} 
\begin{split} 
\int _{ \mathrm{TB} } F ^{ \mathrm{TB}}_\infty \wedge F ^{ \mathrm{TB}}_\infty  &=\int_{\Sigma  _\infty } F ^{ \mathrm{TB}}_\infty  =1,\\
\int _{ \mathrm{TB} } F ^{ \mathrm{TB}}_\infty \wedge F ^{ \mathrm{TB}}_{ \mathrm{bolt}}   &=\int_{ \Sigma  _\infty } F ^{ \mathrm{TB}}_\mathrm{bolt}  
=\int_{ \Sigma  _ \mathrm{b}} F ^{ \mathrm{TB}}_\infty   =0,\\
\int _{ \mathrm{TB} } F ^{ \mathrm{TB}}_{ \mathrm{bolt}} \wedge F ^{ \mathrm{TB}}_{ \mathrm{bolt}}   &
=\int_{ \Sigma  _ \mathrm{b}} F ^{ \mathrm{TB}}_\mathrm{bolt}   =-1.
\end{split} 
\end{equation}
Moreover $ F ^{ \mathrm{TB}}_\infty $, $F ^{ \mathrm{TB}}_{ \mathrm{bolt}} $ are square-integrable as
\begin{equation}
\int _{ \mathrm{TB} } F ^{ \mathrm{TB}}_\infty \wedge *  F ^{ \mathrm{TB}}_\infty = \frac{5}{4} , \qquad 
\int _{ \mathrm{TB} } F ^{ \mathrm{TB}}_ \mathrm{bolt}  \wedge *  F ^{ \mathrm{TB}}_ \mathrm{bolt}   = \frac{5}{4} ,
\end{equation} 
hence they  generate $ L ^2 \mathcal{H} ^2 ( \mathrm{TB} )$.  The 2-form $F ^{ \mathrm{TB}}_{ \mathrm{bolt}} $ is non-exact and generates  $ H ^2 _{ \mathrm{dR}  } (\mathrm{TB}  ) $.
By comparing with (\ref{intforms}), we see that $ F ^{ \mathrm{TB}}_\infty $  
(respectively, $F ^{ \mathrm{TB}}_ \mathrm{bolt} $) is the Poincar\'e dual of $\Sigma  _\infty $ (respectively, $\Sigma  _{ \mathrm{b}} $).

The  (anti) self-dual combination has the simpler expression
\begin{align}
\label{sdtb} 
\FF^{ \mathrm{TB}}   &
= \frac{1}{2}    \left( \mathrm{d} \xi ^\flat   +  * \mathrm{d} \xi ^\flat   \right)  
= - \frac{9}{8} \left(  \frac{e ^1 \wedge e ^2 + e ^3 \wedge e ^4 }{(r + N )^2 } \right)
=\frac{9}{8} \mathrm{d} \left[ \left( \frac{r-N}{r + N } \right) \eta _3 \right] ,\\
\label{asdtb} 
\FFasd^{ \mathrm{TB}}&
=\frac{1}{2} \left( \mathrm{d} \xi ^\flat - * \mathrm{d} \xi ^\flat   \right) 
=\frac{1}{8} \left(  \frac{e ^1 \wedge e ^2 - e ^3 \wedge e ^4 }{(r - N )^2 } \right)
= - \frac{1}{8}  \mathrm{d} \left[ \left( \frac{r + N }{r - N } \right)  \eta _3 \right] .
\end{align} 

Another interesting 2-cycle to consider is the one obtained as we did for $\Sigma _{ \mathrm{c} }$ in ES, that is by considering the circle fibration (except at the endpoints) over a line of constant $\theta$, $\phi$ connecting a point of $\Sigma _ \mathrm{b} $ with a point of $\Sigma _\infty $. We denote it by $\Sigma _ \mathrm{c} $ and equip it with the orientation $\mathrm{d} r \wedge \mathrm{d} \psi  $. One can then check that, as homology classes, $ [ \Sigma _ \mathrm{c} ] =[ \Sigma _\infty ] - [ \Sigma _ \mathrm{b} ] $, and correspondingly the Poincar\'e dual of $\Sigma _ \mathrm{c} $ is $F ^{ \mathrm{TB} }_\infty - F ^{ \mathrm{TB} }_ \mathrm{b}  $.

\subsection{Taub-NUT}
\label{hftn} 
In  the case of TN,
\begin{equation}
\frac{c ^2 }{a ^2 } = - \frac{2 c ^\prime }{f} =\frac{4 N ^2 }{(r + N )^2 },
\end{equation}
hence $\mathrm{d} \xi ^\flat $ is self-dual,
\begin{equation}
\label{dxitn} 
\mathrm{d} \xi ^\flat = * \mathrm{d} \xi ^\flat =\mathrm{d} \left( \frac{r - N }{r+ N } \, \eta _3 \right) = - \left( \frac{e ^1 \wedge e ^2 + e ^3 \wedge e ^4}{(r + N )^2 }\right) .
\end{equation} 
In fact, on a hyperk\"ahler manifold satisfying certain growth conditions, which is the case for TN, any $L ^2 $ harmonic form is necessarily self-dual \cite{Hitchin:398393}.
The 2-form $\mathrm{d} \xi ^\flat $ has been considered several times \cite{Pope:1978ffa,Hitchin:398393,Jante:2014ho,Franchetti:2014ue}. Defining
\begin{equation}
\label{pdiftn} 
F ^{ \mathrm{TN}} _\infty = - \frac{\mathrm{d} \xi ^\flat }{4 \pi } = \frac{1}{4 \pi } \left( \frac{e ^1 \wedge e ^2 + e ^3 \wedge e ^4}{(r + N )^2 } \right) ,
\end{equation}
we have
\begin{equation} 
\label{intftn} 
\int _{ \Sigma  _\infty } F ^{ \mathrm{TN}}_\infty = \int _{ \mathrm{TN}  } F ^{ \mathrm{TN}}_\infty \wedge F ^{ \mathrm{TN}}_\infty =1.
\end{equation} 
Therefore, $F ^{ \mathrm{TN}} _\infty $  is square-integrable and by comparing with (\ref{intforms}) we  see that it
  is the Poincar\'e dual of $\Sigma  _\infty $ in $X _{ \mathrm{TN} }$.
  
We can also consider the 2-cycle $\Sigma _ \mathrm{c}$ obtained by considering the circle bundle (except at the endpoints) over a line of fixed $ \theta $, $\phi$ connecting the nut with a point of $\Sigma _\infty $ and equipped with the orientation $\mathrm{d} r \wedge \mathrm{d} \psi $. One can check that $[ \Sigma _ \mathrm{c} ] = [ \Sigma _\infty ] $.
See  \cite{Franchetti:2014ue} for more details on the harmonic cohomology of TN and multi TN.

\subsection{Euclidean Schwarzschild}
\label{eshfhar} 
In the case of ES we take 
\begin{equation} \xi =\frac{1} {16 M ^2 }  \frac{ \partial}{ \partial \chi }.
\end{equation}
We have
\begin{equation} 
\begin{split} 
\mathrm{d} \xi ^\flat &= \mathrm{d} \left( \left( 1 - \frac{2 M }{r} \right) \mathrm{d} \chi \right)  = \frac{2M}{r ^2 } \mathrm{d} r \wedge \mathrm{d} \chi 
= - \frac{e ^3 \wedge e ^4 }{2r ^2 },\\
* \mathrm{d} \xi^\flat  &=\frac{1}{2} \, \mathrm{d} (\cos \theta \, \mathrm{d} \phi )
= - \frac{1}{2}  \sin \theta \, \mathrm{d} \theta \wedge \mathrm{d} \phi = - \frac{e ^1 \wedge e ^2 }{2r ^2 }.
\end{split} 
\end{equation} 
The form $\mathrm{d} \xi ^\flat $ is exact while $ * \mathrm{d} \xi ^\flat $  is proportional to the area form of the bolt and generates $H ^2 _{ \mathrm{dR}  }(\mathrm{ES} )$.  Defining
\begin{equation}
\label{ffffes} 
F ^{ \mathrm{ES}}_\infty   = - \frac{ \mathrm{d} \xi  ^\flat }{2 \pi }, \qquad F ^{ \mathrm{ES}}_{\Sigma _ \mathrm{c} }   = *F ^{ \mathrm{ES}}_\infty   
= - \frac{* \mathrm{d} \xi ^\flat  }{2 \pi },
\end{equation} 
it can be checked that
\begin{equation}
\label{esvarintoiy} 
\begin{split} 
\int _{ \mathrm{ES} } F ^{ \mathrm{ES}}_\infty \wedge F ^{ \mathrm{ES}}_\infty  &=\int_{ \Sigma  _\infty } F ^{ \mathrm{ES}}_\infty  =0,\\
\int _{ \mathrm{ES} } F ^{ \mathrm{ES}}_\infty \wedge F ^{ \mathrm{ES}}_{\Sigma _ \mathrm{c} }&=\int_{\Sigma _\infty } F ^{ \mathrm{ES}}_{\Sigma _ \mathrm{c} }  
=\int_{\Sigma _ \mathrm{c} } F ^{ \mathrm{ES}}_\infty   =1,\\
\int _{ \mathrm{ES} } F ^{ \mathrm{ES}}_{\Sigma _ \mathrm{c} }  \wedge F ^{ \mathrm{ES}}_{\Sigma _ \mathrm{c} }&=\int_{\Sigma _ \mathrm{c} }F ^{ \mathrm{ES}}_{\Sigma _ \mathrm{c} }=0,
\end{split} 
\end{equation} 
hence $ F ^{ \mathrm{ES}}_\infty $, $ F ^{ \mathrm{ES}}_{\Sigma _ \mathrm{c} } $ generate $L ^2 \mathcal{H} ^2 (\mathrm{ES} )$. By comparing with (\ref{intforms}) we see that they are the Poincar\'e duals of $\Sigma  _\infty $ and $\Sigma_ \mathrm{c} $.
We again have the (anti) self-dual combination 
\begin{equation}
\begin{split} 
\FF^{ \mathrm{ES}}   &
= \frac{1}{2}    \left( \mathrm{d} \xi ^\flat   +  * \mathrm{d} \xi ^\flat   \right)  
= - \frac{1}{4} \,  \frac{  e ^3 \wedge e ^4 +  e ^1 \wedge e ^2 }{ r ^2 }
= - \frac{1}{4 } \left( 
\frac{4M}{r ^2 } \,  \mathrm{d} \chi \wedge \mathrm{d} r + \sin \theta \, \mathrm{d} \theta \wedge \mathrm{d} \phi 
  \right) ,\\
\FFasd^{ \mathrm{ES}}   &
= \frac{1}{2}    \left( \mathrm{d} \xi ^\flat   -  * \mathrm{d} \xi ^\flat   \right)  
= - \frac{1}{4} \,  \frac{  e ^3 \wedge e ^4 -  e ^1 \wedge e ^2 }{ r ^2 }
= - \frac{1}{4 } \left(
\frac{4M}{r ^2 } \,  \mathrm{d} \chi \wedge\mathrm{d} r - \sin \theta \, \mathrm{d} \theta \wedge \mathrm{d} \phi 
 \right).
\end{split} 
\end{equation} 

\newpage 
\section{Harmonic spinors}
\label{hspinors} 
Harmonic spinors are solutions of the Dirac equation $\slashed{ D }  \psi =0 $ with $\psi $ belonging to the 4-dimensional complex spin representation. While TN and ES are spin manifolds, they admit no non-trivial $L ^2 $ harmonic spinors \cite{Pope:1978ffa}; TB is not spin. In order to consider a more interesting problem,  we study the Dirac operator twisted by an Abelian connection having harmonic curvature. For $M$ any of TB, TN, ES, we require our solutions to extend to the HHM compactification $X _M $ of $M$.

Twisting the Dirac operator amounts to tensoring the spinor bundle with a complex line bundle. Complex line bundles over a manifold $X_M$ are classified by  $H^2 (X_M , \mathbb{Z}  )$.  As shown in Section \ref{harforms},  the HHM compactification of TB, TN, ES is homeomorphic to, respectively, $ \mathbb{C}  P ^2 \# \overline{ \mathbb{C}  P }^2 $, $\mathbb{C}  P ^2 $, $ S ^2 \times S ^2 $, hence
\begin{equation}
H^2 (X _{ \mathrm{TB} }, \mathbb{Z}  ) =H ^2 (X _{ \mathrm{ES} }, \mathbb{Z}  ) = \mathbb{Z}  ^2, \quad 
H^2 (X _{ \mathrm{TN} }, \mathbb{Z}  ) = \mathbb{Z}  .
\end{equation} 
Since $S ^2 \times S ^2 $ is spin, the spinor bundle is well defined. A real 2-form $\mathcal{F}$ on ES is the curvature of a connection provided that the cohomology class $ \mathcal{F} /(2 \pi ) $ is integral, resulting in the quantisation conditions (\ref{esquant}) below.

Let  now $X$ be either $ X _{ \mathrm{TB}} $ or $ X _{ \mathrm{TN}} $, which are not spin. Separately, the spinor bundle on $X$ and a line bundle on $X$ with half-integer\footnote{We say that a number $n$ is half-integer if  $n =(2k + 1)/2 $ with $k \in \mathbb{Z}  $.}  Chern number are  not well-defined. However, their tensor product is well defined and equips $X$  with a $\mathrm{Spin} ^ \mathbb{C}  $ structure \cite{Lawson:276634}. Therefore, we require the cohomology class $ \mathcal{F} /2 \pi $ of a real 2-form $\mathcal{F}$  on TB or TN to be half-integral, resulting in the conditions (\ref{tbconds}) for TB and in the condition (\ref{tncond}) for TN.

Zero modes of the twisted Dirac operator on  TN and ES have been studied before \cite{Pope:1978ffa,Jante:2014ho,Jante:2016vq}, although in the case of ES our choice of a connection is more general. TB has not been considered before so we focus on it.

The Dirac operator $\slashed{ D } _\mathcal{A}  $ on a bi-axial Bianchi IX manifold with metric (\ref{b9}) twisted by an Abelian real-valued connection $\mathcal{A}$ has the form, see Appendix \ref{twistdirac},
\begin{equation}
\label{diracmatr} 
\slashed{ D } _\mathcal{A} =
 \begin{pmatrix}
 \mathbf{0}  & \T ^\dagger    \\
\T   & \mathbf{0} 
\end{pmatrix},
\end{equation} 
with
\begin{align} 
\T&=  \frac{i}{f} \left( \partial _r + \frac{a ^\prime }{a} + \frac{c ^\prime }{2c} + \mathcal{A} (E _4  ) \right) \mathbf{1}  + i \dbb,\\
\T ^\dagger  &
= \frac{i}{f} \left( \partial _r  +   \frac{ a ^\prime  }{a} + \frac{ c ^\prime  }{2c} + \mathcal{A} (E _4  )\right) \mathbf{1} - i \dbb,
\end{align} 
\begin{equation}
\dbb  =\frac{1}{2a}\begin{pmatrix}
2i D _3 / \lambda  & 2i D _-  \\
2i D _+  & - 2i D _3 / \lambda 
\end{pmatrix} + \frac{\mathbf{1} }{2a} \left( \frac{2 + \lambda ^2 }{2 \lambda } \right) .
\end{equation} 
Here $\lambda =c/a $, $D _i = X _i + i \mathcal{A} (X   _i)$, $ i =1 ,2,3$, $ D _{ \pm } = D _1 \pm i D _2$.
The vector fields $X _i $ are  dual to the left-invariant 1-forms (\ref{livforms}) and given by
\begin{equation}
\label{xis} 
\begin{split}
X _1 &= \sin \psi\,  \partial _\theta + \frac{\cos \psi }{\sin \theta } \left( \cos \theta \, \partial _\psi - \partial _\phi \right) ,\\
X _2 &= \cos \psi\,  \partial _\theta - \frac{\sin  \psi }{\sin \theta } \left( \cos \theta \, \partial _\psi - \partial _\phi \right) ,\\
X _3 &= \partial _\psi.
\end{split}
\end{equation} 

By the results of Section \ref{harforms}, harmonic connections satisfy $ \mathcal{A} (X _1 )= \mathcal{A} (X _2 ) = \mathcal{A} (\partial _r  ) =0 $, so that $D _\pm  = X _\pm  =X _1 \pm i X _2 $, $D _3 =X _3 + i \mathcal{A} (X _3 )= X _3 + i \p/2 $ with
\begin{equation}
\tilde p =2 \mathcal{A} (X _3 )
\end{equation}
a function of $r$ only. Hence
\begin{equation}
\label{stepxxx} 
\begin{split} 
\T&=  \frac{i}{f} \left( \partial _r + \frac{a ^\prime }{a} + \frac{c ^\prime }{2c} \right) \mathbf{1}  + \frac{i}{2a}\dbbb,    \quad 
\T ^\dagger =  \frac{i}{f} \left( \partial _r + \frac{a ^\prime }{a} + \frac{c ^\prime }{2c} \right) \mathbf{1} - \frac{i}{2a}\dbbb   ,\\
\dbbb &=
 \begin{pmatrix}
(2i X _3- \p) / \lambda  & 2i X _-  \\
2i X _+  & - (2i X _3- \p) / \lambda 
\end{pmatrix} + \left( \frac{2 + \lambda ^2 }{2 \lambda } \right)\mathbf{1}.
\end{split} 
\end{equation} 
The operator $\dbbb $ is essentially the twisted Dirac operator on the squashed 3-sphere, which has been considered in \cite{Pope:1981dj}. A review of its main properties following the notation used here can be found in \cite{Franchetti:2017hs}. We recall in particular that,
because of the spherical symmetry of the problem, the operators $\slashed {D} _\mathcal{A} $,  $\dbbb $  commute with the scalar Laplacian on the round 3-sphere 
\begin{equation}
 \triangle _{ S ^3 } = -(X _1 ^2 + X _2 ^2 + X _3 ^2 ),
 \end{equation} 
hence we can restrict $\dbbb $ to an eigenspace of $ \triangle _{ S ^3 }$.

The eigenspaces of $\triangle _{ S ^3 }$ are given by the irreducible representations $V _j  \otimes V _j  $ of $\mathfrak{ sl } (2, \mathbb{C}  ) \oplus  \mathfrak{ sl }(2, \mathbb{C}  ) $, where $j \geq 0 $, $2j\in \mathbb{Z}  $ and $V _j  $ is the irreducible representation of $\mathfrak{ sl }(2, \mathbb{C}  ) $ of dimension $2j +1 $. We use the shorthand notation $|j,m, m ^\prime \rangle$ for  the element $ |j,m \rangle \otimes |j, m ^\prime \rangle \in V _j \otimes V _j$. Let
\begin{equation} 
\{ |j,m, m ^\prime  \rangle  : m, m ^\prime  \in \{ -j , -j + 1, \ldots , j-1 ,j \}\} 
\end{equation}
be a basis of $V _j \otimes V _j $ consisting of simultaneous eigenvectors of $\triangle _{ S ^3 } $ and $i X _3 $.
It is possible to show that the eigenvectors and eigenvalues of $\dbbb $ are
\begin{alignat}{2}
\label{genegv} 
&\begin{pmatrix}
 C _1  |j,m, m ^\prime  \rangle  \\
 C _2 |j,m+1, m ^\prime  \rangle 
\end{pmatrix} &\quad &
\begin{matrix*}[r]\frac{\lambda}{2} \pm \frac{ 1}{ \lambda } \sqrt{ (2m +1 - \p )^2 + 4 \lambda  ^2 (j-m)(j+m+1) }, \\  -j \leq m \leq j-1
\end{matrix*}
\\   
\label{mjegv} 
&\begin{pmatrix}
0  \\
|j,-j, m ^\prime  \rangle  
\end{pmatrix}
& &
\frac{ \lambda}{2} + \frac{ 1}{\lambda} (2j +1 + \p) , \ m =-j-1 
\\
\label{jegv} 
&\begin{pmatrix}
 |j,j, m ^\prime  \rangle   \\
0 
\end{pmatrix}  &&
\frac{ \lambda}{2} + \frac{ 1}{\lambda} (2j +1 - \p), \ m =j 
\end{alignat}

with  $C _1 / C _2 $  satisfying the relation
\begin{equation}
\label{ccrel} 
\frac{C _1 }{C _2 } 
=\frac{1}{\lambda } \left[ \frac{2m + 1 - \p  \pm  \sqrt{ (2m +1 - \p )^2 + 4 \lambda ^2 (j-m)(j+m+1) }}{2 \sqrt{ (j-m)(j+m+1) }} \right] .
\end{equation} 
All the eigenvectors have multiplicity $2j + 1 $ coming from the allowed values of $m ^\prime \in \{ -j, -j + 1 , \ldots ,j-1, j\} $.

Let us go back to solving the equation $\slashed{D} _\mathcal{A} \psi =0 $.
By  (\ref{diracmatr}), writing
\begin{equation}
 \psi = \begin{pmatrix}
\Psi   \\
\Phi 
\end{pmatrix},
\end{equation} 
with $\Psi, \Phi $ 2-component Weyl spinors,  we get the equations 
\begin{equation}
\T ^\dagger   \Phi =0 =\T  \Psi .
\end{equation}

Consider first the equation $\T \Psi =0 $. Any solution can be written in the form
 \begin{equation}
 \label{psiansatz} 
 \Psi =
h (r) \, \V,
 \end{equation}
with $h$ a radial function  and $ \V $ an eigenvector of  $\dbbb $.
Writing $\Lambda $ for the eigenvalue of $\V$, substituting  in (\ref{stepxxx}) gives 
\begin{equation}
\T \Psi = 
\left[  \frac{i}{f} \left( \partial _r + \frac{a ^\prime }{a} + \frac{c ^\prime }{2c} \right) 
+\frac{i \, \Lambda }{2a}  \right] \Psi 
\end{equation} 
and the equation $\T \Psi =0 $ reduces to the ODE
\begin{equation}
\label{finalode} 
h ^\prime + \left[\frac{a ^\prime }{a} + \frac{c ^\prime }{2c}
+ \frac{f}{2a}\Lambda  \right] h =0,
\end{equation} 
which integrates to
\begin{equation}
\label{sol1aa} 
h =\frac{ h _0}{ a \sqrt{ c }} \, \exp \left( - \frac{1}{2} \int \frac{ f \Lambda }{a} \, \mathrm{d} r \right) ,
\end{equation} 
with $h _0 $ a positive constant.

The equation $ \T ^\dagger  \Phi = 0 $ can be treated similarly by taking
 \begin{equation}
\label{phiansatz}
 \Phi =
k (r) \, \V ,
 \end{equation}
 with $h$ a radial function  and $ \V $ an eigenvector of  $\dbbb $,
to obtain the ODE
\begin{equation}
\label{finalode2} 
k ^\prime + \left[\frac{a ^\prime }{a} + \frac{c ^\prime }{2c}
- \frac{f}{2a}\Lambda  \right] k =0,
\end{equation} 
which integrates to
\begin{equation}
\label{sol2aa} 
k =\frac{ k _0}{ a \sqrt{ c }} \, \exp \left(  \frac{1}{2} \int \frac{ f \Lambda }{a} \, \mathrm{d} r \right) ,
\end{equation} 
with $k _0 $ a positive constant.

We note  that on a manifold with infinite volume and non-negative scalar curvature $s$, $ \T ^\dagger $ (respectively, $ \T$) has no non-trivial $L ^2 $  zero modes if $\mathcal{F} = \mathrm{d} \mathcal{A} $ is self-dual (anti self-dual).
In fact, by the generalised Lichnerowicz's formula, see e.g.~\cite{Lawson:276634},
\begin{equation}
\label{lichntw} 
\slashed{D} _\mathcal{A} ^2  \psi 
= \triangle  \psi + \frac{s}{4} \psi + \frac{1}{2} \mathcal{F}  _{ \mu \nu }\gamma _\mu \gamma _\nu    \psi,
\end{equation} 
 where  $ \triangle = \nabla \circ \nabla $ is the connection Laplacian and $ \nabla $ the covariant derivative associated to the spin connection.
If $\mathcal{F}$ is self-dual (anti self-dual), the last term of (\ref{lichntw}) vanishes when $\psi$ is a right-handed (left-handed) spinor. In either case since $s \geq 0 $, using the fact that $ \slashed{D} _\mathcal{A}$, $\nabla $ are formally self-adjoint, we obtain
\begin{equation}
0= \langle \nabla \psi , \nabla \psi \rangle + \frac{s}{4} \langle  \psi , \psi \rangle ,
\end{equation} 
so that a square-integrable spinor $\psi $ has to vanish or, if $s =0 $, to be covariantly constant, hence zero in a manifold with infinite volume.

 In the next subsections we specialise to TB, TN and ES with the Dirac operator twisted by a general (i.e.~not necessarily self-dual) harmonic connection.

\subsection{Taub-bolt}
\label{khjadsrewoui} 
As we have seen in Section \ref{hftb}, a generic harmonic $L ^2 $ 2-form $\mathcal{F}$  on TB can be written 
\begin{equation}
\label{tnfs} 
\mathcal{F} = -  \pi   \left( A  F _\infty ^ \mathrm{TB} + B F _ \mathrm{bolt} ^ \mathrm{TB} \right) 
=\left(   \frac{3 A + 5 B }{12} \right)   \mathrm{d} \xi ^\flat - \frac{B }{3} \, * \mathrm{d} \xi ^\flat ,
\end{equation} 
with $A ,B $ constants. It can be checked that $\mathcal{F} =\mathrm{d} \mathcal{A}  $ with
\begin{equation}
\label{atb} 
\mathcal{A} =  \left( \frac{3 A + 5 B }{12}\,   U  -  \frac{B }{6N}  ( r ^2 - N ^2 )U ^\prime  \right)  \eta _3  ,
\end{equation} 
with  $U$ given by (\ref{ufunct}).
Unless $B=0 $,  $\mathcal{A}$ is not well-defined on the bolt $r =2 N $ since $\eta _3 $ is not defined there. 
The 2-form $\mathcal{F} $ is globally defined but not exact.

As discussed at the beginning of Section \ref{hspinors},  $ \mathcal{F} / ( 2 \pi )$ needs to be a half-integral cohomology class, hence we impose the conditions
\begin{equation} 
\label{tbconds} 
-  \frac{1}{2 \pi } \int _{\Sigma _ \mathrm{b}   } \mathcal{F}   
=  p + \frac{1}{2} , \quad p \in \mathbb{Z}, \quad 
- \frac{1}{2 \pi } \int _{ \Sigma  _\infty } \mathcal{F} 
= q + \frac{1}{2} , \quad q \in \mathbb{Z}  .
\end{equation} 
By (\ref{tbvarint}),
\begin{equation}
-  \frac{1}{2 \pi } \int _{ \Sigma _ \mathrm{b}   } \mathcal{F}  
=- \frac{B}{2}  ,\quad 
- \frac{1}{2 \pi } \int _{ \Sigma _\infty } \mathcal{F}
=\frac{A }{2},
\end{equation} 
hence we find the conditions
\begin{equation}
\label{abval} 
B = - (2p +1 )  , \quad A = 2q + 1 , \quad p,q \in \mathbb{Z} 
\end{equation} 
and (\ref{tnfs}) becomes
\begin{equation}
\label{tbtbtbffff} 
\mathcal{F} =-  \pi \Big( ( 2q + 1) F ^{ \mathrm{TB} }_\infty  - (2p + 1) F ^{ \mathrm{TB} }_ \mathrm{bolt}   \Big)
= \left( \frac{3q-5p-1}{6} \right) \mathrm{d}  \xi ^\flat + \left( \frac{2 p + 1 }{3} \right) * \mathrm{d} \xi ^\flat   .
\end{equation}

It is shown in Appendix \ref{othereigen} that taking $\V$ in (\ref{sol1aa}) or (\ref{phiansatz})  to be an eigenvector of $\dbbb $ of the form (\ref{genegv}) leads to no non-trivial $L ^2 $ harmonic spinors. Here we consider the choices (\ref{jegv}), for which $ \upsilon = (|j , m= j , m ^\prime \rangle ,0) ^T  $, and (\ref{mjegv}), for which $ \upsilon = (0,|j ,  m=-j , m ^\prime \rangle ,0) ^T  $. For the sake of brevity, in the following  the cases (\ref{jegv}), (\ref{mjegv}) will be referred to as $m =j $, $m =-j $ respectively.

Using the values in (\ref{tbvalues}), we calculate
\begin{equation}
\frac{a _{ \mathrm{TB}}^\prime }{a_{ \mathrm{TB}}} + \frac{c _{ \mathrm{TB}}^\prime }{2c_{ \mathrm{TB}}} = \frac{8 r ^3 -15 r ^2 N + 5 N ^3 }{8 a _{ \mathrm{TB}}^2 V }, \quad V =(r - 2 N )(r - N/2) =\left(  \frac{c _{ \mathrm{TB}}a_{ \mathrm{TB}} }{2N}\right) ^2 .
\end{equation}

Consider first the case $m =j $.
Substituting the corresponding eigenvalue for $\Lambda$ in (\ref{finalode}) we get the ODE 
\begin{equation}
\begin{split}
-2 h ^\prime  + \left( 
- \frac{8 r ^3 - 15 r ^2 N + 5 N ^3 }{4 a_{ \mathrm{TB}} ^2 V } + (2j +1 - \p)\frac{ a _{ \mathrm{TB}}^2 }{2 N V } + \frac{N }{a_{ \mathrm{TB} }^2 }
\right)  h =0,
\end{split}
\end{equation} 
with 
\begin{equation}
\label{ptiltb} 
\tilde p =2 \mathcal{A} (X _3 ) =  \frac{3 A + 5 B }{12}\,   U  -  \frac{B }{6N}  ( r ^2 - N ^2 )U ^\prime   .
\end{equation}
The ODE has solution
\begin{equation}
\label{sol} 
h=\frac{ C}{\sqrt{ r + N }}\,  \mathrm{e} ^{ \tfrac{(1 + 4j - 2q )r }{8N}} (r- 2 N )^{ j - \tfrac{p }{2} } ( r - N/2 ) ^{ \tfrac{4j-1 + 2 p }{16}},
\end{equation} 
with $C$ a constant.

The case $m =-j $,  can be obtained replacing $\p $ by $-\p $, or equivalently $q + \tfrac{1}{2} $ by $- \left( q+ \tfrac{1}{2} \right) $ and $p +  \tfrac{1}{2} $ by $- \left( p+ \tfrac{1}{2} \right) $.  The solution of the corresponding ODE is
\begin{equation}
\label{solmneg} 
\begin{split} 
h &=\frac{ C}{\sqrt{ r + N }}\,  \mathrm{e} ^{ \tfrac{(4j + 2q + 3 )r }{8N}} (r- 2 N )^{  \tfrac{1 + 2j + p }{2} } ( r - N/2 ) ^{ \tfrac{4j-3 - 2 p }{16}}.
\end{split} 
\end{equation} 

The $m =j $ and $m =-j $ solutions can be written in a unified format as
\begin{equation}
\label{tbsoluni} 
h = \frac{C }{\sqrt{ r + N }} \, 
\mathrm{e} ^{
 \frac{ \left(  2j+1 \mp \left ( q + \frac{1}{2} \right ) \right) r}{4 N } } 
\left( r - 2 N \right) ^{ j + \frac{1}{4} \mp \frac{1}{2} \left ( p + \frac{1}{2} \right )}  
\left( r -  N /2 \right) ^{ \frac{j}{4} - \frac{1}{8} \pm \frac{1}{8}\left ( p + \frac{1}{2} \right ) },
\end{equation} 
with the top (bottom) sign if $m =j$ ($m =-j $).

The volume form 
\begin{equation}
- a_{ \mathrm{TB}} ^2 \, c_{ \mathrm{TB}} \, f_{ \mathrm{TB}} \mathrm{d} r \wedge \eta _1 \wedge \eta _2 \wedge \eta _3 
=   2 N (r ^2- N ^2 ) \mathrm{d} r \wedge \eta _1 \wedge \eta _2 \wedge \eta _3 
\end{equation} 
 does not affect square-integrability at either $r =2 N $ or $\infty $.
The solution (\ref{tbsoluni}) is $L ^2 $  around  $ r =2 N $ if and only if 
$2j \mp \left( p + \tfrac{1}{2} \right)  + \tfrac{1}{2} > - 1$. Since $ 2j,p\in \mathbb{Z}   $, equivalently
\begin{equation}
\label{hcondr} 
2j  + 1 \geq  \pm \left( p + \frac{1}{2} \right) + \frac{1}{2}  .
\end{equation} 
Square-integrability at large $r $  gives
$ 2j +1 \mp  \left ( q + \tfrac{1}{2} \right ) <0$,
or 
\begin{equation}
\label{condd2} 
 2j + 1 \leq \pm \left ( q + \frac{1}{2}  \right ) - \frac{1}{2}  .
\end{equation} 
The two conditions taken together read
\begin{equation}
\label{unicondh} 
\pm \left ( p + \frac{1}{2} \right )  + \frac{1}{2} \leq 2j + 1 \leq  \pm\left ( q + \frac{1}{2} \right ) - \frac{1}{2}.
\end{equation} 
Hence $q \geq 2j + 1 \geq 1 $, $p \leq q-1  $ if $m =j $;  $q \leq -2j -2 \leq -2 $, $p \geq q + 1  $ if $m = - j $.

The case $\T ^\dagger \Phi =0 $ can be treated similarly. Without giving the full details, we find
\begin{equation}
\label{ksolllljk} 
k = \frac{C }{\sqrt{ r + N }} 
\mathrm{e} ^{
 \frac{ - \left(  2j+1 \mp\left ( q + \frac{1}{2} \right ) \right) r}{4N } } 
\left( r - 2 N \right) ^{ - \left(  j +  \frac{3}{4} \mp\frac{1}{2} \left ( p + \frac{1}{2} \right ) \right) }  
\left( r -  N /2 \right) ^{ - \left( \frac{j}{4} + \frac{3}{8} \mp\frac{1}{8} \left ( p + \frac{1}{2} \right )\right)},
\end{equation}
with the top (bottom) sign if  $m =j $ ($m =-j$).
Square integrability at $r =2 N $  gives $2j \mp \left( p + \tfrac{1}{2} \right) < - 1/2$, or equivalently
\begin{equation}
\label{kcondr} 
2j \mp \left ( p + \frac{1}{2} \right ) \leq  - \frac{3}{2}.
\end{equation} 
Square integrability at large $r$  gives
$2j + 1 \mp\left ( q + \tfrac{1}{2} \right )  >0 $,
or 
\begin{equation}
2j + 1 \geq \pm \left ( q + \frac{1}{2} \right ) + \frac{1}{2}.
\end{equation} 
The two conditions taken together read
\begin{equation}
\label{unicondk} 
\pm\left ( q + \frac{1}{2} \right ) + \frac{1}{2} \leq 2j + 1 \leq \pm\left ( p + \frac{1}{2} \right ) - \frac{1}{2} .
\end{equation}
Hence $p \geq 2j + 1 \geq 1 $, $q \leq p -1 $ if $m =j $; $p \leq -2j-2 $, $q \geq p + 1 $ if $m =- j $. Note that (\ref{unicondk}) is equal to (\ref{unicondh}) with $p \leftrightarrow q $.
By comparing (\ref{unicondh}) with (\ref{unicondk}) we see that if  $\T ^\dagger $  admits non-trivial $L ^2 $  zero modes then $\T  $  does not, and viceversa. 

Each zero mode has multiplicity $2j+1 $, so if $\pm\left ( p + \frac{1}{2} \right )  + \frac{1}{2} \leq \pm\left ( q + \frac{1}{2} \right ) - \frac{1}{2} $ then $\T $ has 
\begin{equation}
\label{genzero} 
\sum _{ i =\pm\left ( p + \tfrac{1}{2} \right ) + \tfrac{1}{2} }^{ \pm\left (  q + \tfrac{1}{2} \right ) - \tfrac{1}{2} } i
=\frac{ q (q + 1 )}{2}- \frac{ p (p + 1 )}{2}
\end{equation} 
$L ^2 $ zero modes of the form (\ref{tbsoluni}). If $\pm\left ( q + \frac{1}{2} \right ) + \frac{1}{2} \leq \pm\left ( p + \frac{1}{2} \right ) - \frac{1}{2} $ then $\T ^\dagger $ has 
\begin{equation}
\label{genzerod} 
- \left( \frac{q (q + 1 )}{2}-\frac{ p (p + 1 )}{2} \right) 
\end{equation}
 $L ^2 $ zero modes of the form (\ref{ksolllljk}).

\subsubsection*{Self-dual case}
By (\ref{tbtbtbffff}), $\mathcal{F}$ is self-dual for $  q =3p + 1 $.  In this case, as discussed at the end of Section \ref{hspinors}, $\T ^\dagger $ has no non-trivial $L ^2 $ zero modes.
The solution (\ref{tbsoluni}) reduces to
\begin{equation}
\label{sdansatz} 
h = \frac{C }{\sqrt{ r + N }} 
\mathrm{e} ^{
 \frac{ \left(  2j+1 \mp 3\left ( p + \frac{1}{2} \right ) \right) r}{4 N } } 
\left( r - 2 N \right) ^{ j + \frac{1}{4} \mp \frac{1}{2} \left ( p + \frac{1}{2} \right )}  
\left( r -  N /2 \right) ^{ \frac{j}{4} - \frac{1}{8} \pm \frac{1}{8}\left ( p + \frac{1}{2} \right ) } .
\end{equation} 
Condition (\ref{unicondh}) becomes
\begin{equation}
\pm\left ( p + \frac{1}{2} \right )  + \frac{1}{2} \leq 2j + 1 \leq \pm3 \left ( p + \frac{1}{2} \right ) - \frac{1}{2} .
\end{equation} 
In particular, $p \geq 0 $ for $m =j $ and $p \leq -1 $ for $m =-j $.
By (\ref{genzero}),  for any $p \in \mathbb{Z}  $ the number of $L ^2 $ harmonic spinors is
\begin{equation}
\label{spinorcount} 
\sum _{ i =\pm\left ( p + \tfrac{1}{2} \right ) + \tfrac{1}{2} }^{ 3 \left( \pm\left (  p + \tfrac{1}{2} \right ) + \frac{1}{2} \right) - 2 } i
=(2p + 1 )^2 .
\end{equation} 

\subsection{Taub-NUT}
Harmonic spinors on  TN have been studied in \cite{Pope:1978ffa,Jante:2014ho,Franchetti:2017hs} so we only recall briefly the relevant results.  As discussed in Section \ref{hftn}, any harmonic $L ^2 $ form on TN is self-dual and  can be written
\begin{equation}
\label{ftnhjg} 
\mathcal{F} = -2 \pi A \, F ^ \mathrm{TN} _\infty = \frac{A}{2} \, \mathrm{d} \xi ^\flat = \mathrm{d} \mathcal{A} ,
\end{equation}
 with $A$ a constant and
\begin{equation}
\label{tnconnect} 
\mathcal{A} =\frac{A}{2} \left( \frac{r- N }{r + N }\right) \eta _3.
\end{equation}
In order for the cohomology class $\mathcal{F} /( 2 \pi ) $ to be half-integral we require
\begin{equation} 
\label{tncond} 
- \frac{1}{2 \pi } \int _{ \Sigma _\infty } \mathcal{F} =A =q + \frac{1}{2}, \quad q \in \mathbb{Z},
\end{equation} 
hence
\begin{equation}
\mathcal{F} = - \pi (2q + 1 )F ^{ \mathrm{TN} }_\infty .
\end{equation}

Since $\mathcal{F} $ is self-dual, only $\T $ has non-trivial $L ^2 $ zero modes. It can be shown \cite{Jante:2014ho,Franchetti:2017hs} that these zero modes are of the form 
\begin{equation}
 \Psi= 
 \begin{cases}
 \begin{pmatrix}
K _1  \\
0
\end{pmatrix}
h  (r) \, |j,j, m ^\prime  \rangle  &\text{for $q \geq 0 $},\\
 \begin{pmatrix}
0  \\
K _2
\end{pmatrix}
h  (r) \, |j, -j, m ^\prime  \rangle  &\text{for $q <0 $}.
\end{cases} 
\end{equation} 
The radial function $h  $  satisfies a different ODE. In terms of the coordinates used here,  $h$  is given by  \cite{Jante:2014ho} 
%\footnote{In  \cite{Jante:2014ho} the TN metric is given in coordinates $(u, \theta , \phi , \psi )$  such that
%\begin{equation*}
%g=  V (\mathrm{d} u ^2 + u ^2 \mathrm{d} \Omega ^2 ) + L ^2 V ^{-1} \eta _3 ^2, \quad V =1 + \frac{L}{u}.
%\end{equation*} 
%The coordinates $(\theta , \phi,  \psi) $  are equal to those used here  in (\ref{tnm})  and have the same range, while $u \in [0, \infty )$. The coordinate transformation $r = u + L /2 $ followed by the redefinition $N =L /2 $ makes $g$ equal to (\ref{tnm}).}
\begin{equation}
\label{tnsolhar} 
h = C\, \frac{( r - N )^j }{\sqrt{ r + N }} \mathrm{e} ^{ \frac{ \left( 2j + 1 \mp \left ( q + \frac{1}{2}  \right ) \right) r}{4 N } },
\end{equation} 
with $C$ an arbitrary constant and the top (bottom) sign if $ m =j $ ($m =-j $). It is $L ^2 $ for $  2j + 1 < \pm\left ( q + \tfrac{1}{2} \right ) $, or equivalently 
\begin{equation}
\label{tnrangesu2} 
1 \leq 2j + 1 \leq \pm\left ( q + \tfrac{1}{2} \right ) - \tfrac{1}{2} .
\end{equation}
 The number of $L ^2 $  harmonic spinors is
\begin{equation}
\label{tnhs} 
\sum _{ i =1 }^{ \pm\left ( q + \tfrac{1}{2} \right ) - \tfrac{1}{2} }i = \frac{q (q + 1 )}{2}.
\end{equation} 

\subsection{Euclidean-Schwarzschild}
\label{eszm} 
Harmonic spinors on ES have been studied in \cite{Jante:2016vq}, where, however, only the case of a self-dual $L ^2 $ connection is considered. We treat here the more general case of a generic $L ^2 $ harmonic connection.

As discussed in Section \ref{eshfhar}, any $L ^2 $ harmonic form on ES can be written
\begin{equation}
\label{escurvv} 
\mathcal{F} =- 2 \pi \left( q\,  F ^{ \mathrm{ES}} _\infty  + p \,  F ^{ \mathrm{ES}}_{ \Sigma _ \mathrm{c}} \right)=p \, * \mathrm{d} \xi ^\flat + q\,  \mathrm{d} \xi ^\flat = \mathrm{d} \mathcal{A} ,
\end{equation} 
with
\begin{equation}
\label{esconn} 
 \mathcal{A} = \frac{p}{2}\eta _3 + q \left( 1- \frac{2 M }{r} \right) \mathrm{d} \chi .
\end{equation} 
In order for $\mathcal{F}$ to be the curvature of a connection on  $X _{ \mathrm{ES} }$, we require
\begin{equation}
\label{esquant} 
- \frac{1}{2 \pi } \int _{ \Sigma _ \mathrm{c}  } \mathcal{F}  = q\in \mathbb{Z}  , \quad 
- \frac{1}{2 \pi } \int _{ \Sigma _ \infty  } \mathcal{F}  =p \in \mathbb{Z}  .
\end{equation}

Since ES is not of Bianchi IX form, the treatment in Section \ref{hspinors} needs to be modified. We follow \cite{Jante:2016vq} but for a few changes in the notation and the more general choice of $\mathcal{A} $.

We take orthonormal coframe
\begin{equation}
e _1 = a _{ \mathrm{ES}  } \, \mathrm{d} \theta , \quad 
e _2 = a _{ \mathrm{ES}  } \sin \theta\,  \mathrm{d} \phi , \quad 
e _3  = c _{ \mathrm{ES}  } \, \mathrm{d} \chi , \quad 
e _4 = - f _{ \mathrm{ES}  }\,  \mathrm{d} r.
\end{equation} 
For the remaining of this section we drop the  ES decoration in $a,c,f $.
The associated  spin connection is
\begin{equation}
\begin{split}
\omega _{ 14 }&= - \frac{a ^\prime }{af} e ^1 , \quad \omega _{ 24 }= - \frac{a ^\prime }{af} e ^2 , \quad
 \omega _{ 34 }= - \frac{c ^\prime }{cf} e ^3 , \quad 
\omega _{ 12 }= - \frac{\cos \theta }{a \sin \theta } e ^2 , \quad \omega _{ 23 }=\omega _{ 13 }=0.
\end{split} 
\end{equation} 
The twisted Dirac operator (\ref{sled}) is
\begin{equation}
\slashed{ D } _\mathcal{A} =
 \begin{pmatrix}
 \mathbf{0}  & \widetilde {\T ^\dagger}    \\
\widetilde {\T}   & \mathbf{0} 
\end{pmatrix}
\end{equation} 
with 
\begin{align}
 \widetilde \T& =  - \begin{pmatrix}
 -\frac{i}{f} \left( \partial _r + \frac{a ^\prime }{a} + \frac{c ^\prime }{2c} \right)+\frac{1}{c} (\partial _\chi + i\mathcal{A} (\partial _ \chi  )) &
\frac{1}{a}\left(  \partial _\theta - \frac{i}{\sin \theta } \partial _\phi + \tilde s \cot \theta \right)   \\
\frac{1}{a}\left(  \partial _\theta + \frac{i}{\sin \theta } \partial _\phi -s \cot \theta \right)   & 
- \frac{i}{f} \left( \partial _r + \frac{a ^\prime }{a} - \frac{c ^\prime }{2c} \right) - \frac{1}{c} (\partial _\chi + i \mathcal{A} (\partial _ \chi  ))
\end{pmatrix}  ,\\
\widetilde{ \T ^\dagger} & = \begin{pmatrix}
  \frac{i}{f} \left( \partial _r + \frac{a ^\prime }{a} + \frac{c ^\prime }{2c} \right)   +  \frac{1}{c} (\partial _\chi + i\mathcal{A} (\partial _ \chi  )) & \frac{1}{a}\left(  \partial _\theta - \frac{i}{\sin \theta } \partial _\phi + \tilde s \cot \theta \right)   \\
\frac{1}{a}\left(  \partial _\theta + \frac{i}{\sin \theta } \partial _\phi -s \cot \theta \right)   &
   \frac{i}{f} \left( \partial _r + \frac{a ^\prime }{a} + \frac{c ^\prime }{2c} \right)   - \frac{1}{c} (\partial _\chi +i \mathcal{A} (\partial _ \chi  ))
\end{pmatrix}  ,
\end{align} 
where we have defined
\begin{equation}
s =\frac{p-1}{2}, \quad \tilde s =\frac{p + 1}{2}.
\end{equation} 

It can be shown \cite{Jante:2016vq} that the gauge transformation  $ \slashed{D} _\mathcal{A} \rightarrow G \slashed{D} _\mathcal{A} G ^{-1} $ with 
\begin{equation} 
 G =\mathrm{diag}  \big(\exp(is \phi ),\exp(i \tilde s \phi ),\exp(is \phi ),\exp(i \tilde s \phi ) \big)
 \end{equation}
%  leaves $(\T) _{ ii }, (\T ^\dagger) _{ ii }$ unaffected and maps 
%\begin{equation}
%\begin{split} 
% (\T) _{ 21 }&\mapsto \exp(i \tilde s \phi )(\T) _{ 21 }\exp(-is \phi ) , \quad 
%(\T ^\dagger) _{ 21 }\mapsto \exp(i \tilde s \phi )(\T ^\dagger) _{ 21 }\exp(-is \phi ),\\
% (\T) _{12}&\mapsto \exp(i  s \phi )(\T) _{12}\exp(- \tilde is \phi ), \quad 
% (\T ^\dagger) _{12} \mapsto \exp(i  s \phi )(\T^\dagger ) _{12}\exp(- \tilde is \phi ) .
%\end{split}
%\end{equation}
%Using the relations
%\begin{equation}
%\begin{split}
%\exp(i \tilde s \phi )\eth_s  \exp(-i s \phi )&=(1 + |z |^2) \partial _{ \bar z } + s z =-i X _+ ,\\
%\exp(i \tilde s \phi )\overline{ \eth} _{ \tilde s } \exp(-i s \phi )&=(1 + |z |^2) \partial _{ \bar z } - \tilde s \bar z = + i X _- ,
%\end{split}
%\end{equation} 
 maps $\widetilde {\T} $, $\widetilde{ \T} ^\dagger $ to $\T $, $\T ^\dagger $, given by
\begin{align}
  {\T}& 
= \frac{i}{f} \left( \partial _r + \frac{a ^\prime }{a} + \frac{c ^\prime }{2c} \right) \mathbf{1} 
- \frac{1}{c} (\partial _\chi + i\mathcal{A} (\partial _ \chi  )) \sigma _3  - \frac{i}{a} \slashed{D} _{ S ^2, p } , \\
 \T ^\dagger & 
=   \frac{i}{f} \left( \partial _r + \frac{a ^\prime }{a} + \frac{c ^\prime }{2c} \right) \mathbf{1} 
+ \frac{1}{c} (\partial _\chi + i\mathcal{A} (\partial _ \chi  )) \sigma _3 + \frac{i}{a} \slashed{D} _{ S ^2, p } ,
\end{align} 
where $ \slashed{D} _{ S ^2, p } $ is the  Dirac operator on $S ^2 $ twisted by the line bundle with Chern number $p$. We refer to \cite{Jante:2014ho} for  more details on $\slashed{D} _{ S ^2 ,p } $ and its  zero modes. They belong to the $SU (2) $ representation of dimension $|p |$.  We denote by  $\hsp$ ($\hsm $)  a zero mode of $\slashed{D} _{ S ^2 ,p } $ with $p \geq 1 $ ($p \leq -1 $) and recall that   only the top (bottom) component of $\hsp $ ($\hsm$)  is non-zero.
%recall that  zero modes of $ \slashed{D} _{ S ^2 ,p } $ have the form (\red{pulled back to $U _N $})
%\begin{equation}
%\hsp=
%\begin{pmatrix}
% (1 + |z |^2 ) ^{ -s } p _1 (z)  \\
% 0
%\end{pmatrix} 
%\end{equation} 
%for $p \geq 1 $, and
%\begin{equation}
%\hsm=
%\begin{pmatrix}
%0 \\
% (1 + |z |^2 ) ^{ \tilde s } p _2 (\bar z) 
%\end{pmatrix} 
%\end{equation} 
%for $p \leq -1 $. Here $p _1 $ is a polynomial of degree $\leq p-1 $ and $p _2 $ is a polynomial of degree  $\leq -p-1 $. 

%For $\mathcal{A} $ self-dual ($p =\tilde p $), a Licherowicz's argument shows that only left-handed harmonic $L ^2 $ spinors can be non-trivial.
Writing $\psi = ( \Psi , \Phi )^T $ the Dirac equation gives $\T \Psi =0 =\T ^\dagger \Phi $.
Consider first the equation $\T  \Psi =0 $.
We take the ansatz 
\begin{equation}
\label{haresansatz} 
\Psi _{ \pm }=
\hspm h _{ \pm }(r) \mathrm{e} ^{ \pm i \left( n + \tfrac{1}{2}\right)  \chi },
\end{equation} 
with $n \in \mathbb{Z}  $ and  the upper (lower) sign for $p \geq 1 $ ($p \leq -1 $).
 Using the relations
\begin{equation}
\begin{split}
 \frac{f}{c}  \left( 1- \frac{2 M }{a} \right)  =- \frac{a ^\prime }{4 M }, \quad 
 \frac{f}{c}  =- \frac{a ^\prime }{4 M } - \frac{1}{2} (\log a c ^2 ) ^\prime ,
\end{split}
\end{equation}
the equation $\T \Psi =0 $  reduces to the ODE
\begin{equation}
(\log h_\pm) ^\prime  + \log (a \sqrt{ c }) ^\prime =\pm \left( n + \frac{1}{2} \right) (\log c \sqrt{ a })^\prime
\pm   \left( n + \frac{1}{2} -  q \right) \frac{a ^\prime }{4 M }
\end{equation} 
which has solution
\begin{equation}
\begin{split} 
\label{harmes} 
h _\pm&
=\widetilde C_\pm \,  c ^n  a ^{ \frac{n}{2} - \frac{3}{4} } \mathrm{e} ^{  \left( n + \frac{1}{2} \mp q \right) \frac{a}{4 M }  }=C_\pm \,  \left( 1 - \frac{2 M }{r} \right)  ^{n/2}  r ^{ \frac{n}{2} - \frac{3}{4} } \mathrm{e} ^{ \left( n + \frac{1}{2} \mp q \right) \frac{r}{4 M }},
\end{split} 
\end{equation} 
with $\widetilde C_\pm$ a constant, $C _\pm = (4 M )^n \widetilde C _\pm $. The solution (\ref{harmes})  is $L ^2  $ for 
\begin{equation}
\label{esl21} 
 0 \leq n \leq \pm q -1 .
 \end{equation}
Since the zero mode $\hspm$ of  $\slashed{D} _{ S ^2 ,p } $ has multiplicity $|p|$,  for either sign choice the space of $L ^2 $  zero modes of $\T $  obtained from the ansatz (\ref{haresansatz})  has dimension 
\begin{equation}
\label{esdim1} 
pq.
\end{equation}

Consider now  $\T ^\dagger  \Phi =0 $ and  take the ansatz 
\begin{equation}
\label{anses2} 
\Phi _\pm=
\hspm k_\pm (r) \mathrm{e} ^{\pm i \left( n + \tfrac{1}{2}\right)  \chi }
\end{equation} 
with $n \in \mathbb{Z}  $ and the upper (lower) sign if $p \geq 1 $ ($p \leq -1 $). The equation $\T ^\dagger \Phi =0 $ reduces to the ODE
\begin{equation}
(\log k_\pm) ^\prime  + \log (a \sqrt{ c }) ^\prime = \pm\left( n + \frac{1}{2} \right) (\log c \sqrt{ a })^\prime \pm \left( n + \frac{1}{2} + q \right) \frac{a ^\prime }{4 M },
\end{equation} 
which has solution
\begin{equation}
\label{esharmsol2} 
\begin{split} 
k_\pm &=\widetilde K _{ \pm} \,  c ^n  a ^{ \frac{n}{2} - \frac{3}{4} } \mathrm{e} ^{ \left( n + \frac{1}{2} \pm q\right) \frac{a}{4 M }}
=K_\pm \,  \left( 1 - \frac{2 M }{r} \right)  ^{n/2}  r ^{ \frac{n}{2} - \frac{3}{4} } \mathrm{e} ^{ \left( n + \frac{1}{2} \pm q \right) \frac{r}{4 M } },
\end{split} 
\end{equation} 
with $\widetilde K_\pm$ a constant, $K _\pm =(4 M )^n \widetilde K _\pm $. The solution (\ref{esharmsol2}) is $L ^2  $ for 
\begin{equation}
\label{esl22} 
 0 \leq n \leq \mp q-1 ,
 \end{equation}
hence for either sign choice the  space of $L ^2 $ zero modes of $\T ^\dagger $ coming from the ansatz (\ref{anses2}) has dimension 
 \begin{equation}
 \label{esdim2} 
-pq.
 \end{equation}

By (\ref{escurvv}), $\mathcal{F}$ is self-dual if $p =q $, in which case there are exactly $|p |^2 $ square-integrable zero modes.
By comparing the $L ^2 $ conditions (\ref{esl21}), (\ref{esl22}) we see that if $\T $ admits non-trivial $L ^2 $ zero modes, $\T ^\dagger $ does not, and viceversa.

\newpage 
\section{The index of $\slashed{D} _\mathcal{A} $}
\label{apsthmasf} 
If $(M,g)$ is a Riemannian oriented 4-manifold with boundary $\partial M $ and $\slashed{D} _\mathcal{A}  $ is the Dirac operator twisted by a connection $\mathcal{A}$ with curvature $\mathcal{F}$, by the Atiyah-Patodi-Singer (APS) index theorem, see e.g.~\cite{Eguchi:479050},
\begin{equation}
\label{ithm} 
\begin{split} 
\mathrm{index} (\slashed{D} _{ \mathcal{A} })&
= \mathrm{dim} (\mathrm{Ker} (\T) ) -  \mathrm{dim} (\mathrm{Ker} (\T ^\dagger )  ) \\ &
=    \frac{1}{192 \pi ^2 } \int _{M}  \mathrm{Tr} (\Omega ^2 )  
+ \frac{1}{8 \pi ^2 }\int _{M}  \mathcal{F} \wedge \mathcal{F}  
- \frac{1}{192 \pi ^2 } \int _ { \partial M } \mathrm{Tr} \left( \theta \wedge \Omega  \right) 
- \frac{1}{2} (\eta (0) + h).
\end{split} 
\end{equation} 
Here  $\Omega$ is the curvature of some connection on $M$, $\mathrm{Tr}  (\Omega ^2 )= - \Omega _{ ab } \wedge \Omega _{ ab } $, $\theta $ is the second fundamental form of $\partial M $ and  $ \eta (0) + h $ is a non-local boundary contribution depending on the spectrum of the  Dirac operator induced on the boundary.

The first two terms are the usual bulk contributions to the index, the third one is a local boundary contribution. The fourth term is a non-local boundary term related to the spectrum of the Dirac operator induced on the boundary. The operator $\slashed{D} _{ \mathcal{A} } $ in (\ref{ithm}) acts on $L ^2 $ spinors satisfying certain global boundary conditions which are trivially satisfied in the case of $L ^2 $ spinor on a non-compact manifold having infinite volume.

The third  term  only depends on the asymptotic geometry of the manifold and vanishes if $ \partial M $ has the structure of a metric product  \cite{Gilkey:1993dm}, as is the case for ES. An explicit computation shows that it vanishes on TN \cite{Pope:1978ffa} and hence also on TB which has the same asymptotic geometry. 

The fourth term only depends on the asymptotic  behaviour of $M$ and  $\mathcal{A}$. It vanishes on ES \cite{Pope:1978ffa}. In  \cite{Pope:1981dj} it has been computed for a TN-like asymptotic and a connection  with asymptotic form 
\begin{equation}
\mathcal{A} \simeq \frac{\ell}{2} \eta _3 
\end{equation} 
with $\ell$ a constant, with the result
\begin{equation}
\eta (0) + h = - \frac{1}{6} + \ell ^2 - [|\ell|]([|\ell|]+1),
\end{equation} 
where $[x] $ is the largest integer strictly smaller than $x$.

\subsection{Taub-bolt}
From (\ref{atb}), (\ref{abval})  we find 
$\ell =q + \tfrac{1}{2}  $, $q \in \mathbb{Z}  $, so that $ [ |\ell |] = q $ if $q \geq 0 $ and $[ |\ell |] =-q -1 $ if $q \leq -1 $. In either case
\begin{equation}
\eta (0) + h 
=  \frac{1}{12} .
\end{equation} 
Using (\ref{tbtbtbffff}), (\ref{tbvarint})  we calculate
\begin{equation}
\frac{1}{8 \pi ^2 }\int _{ \mathrm{TB} } \mathcal{F} \wedge  \mathcal{F} = \frac{ q (q + 1 ) - p (p + 1 )}{2}.
\end{equation} 
Using e.g.~the curvature form $\Omega$ of  Levi-Civita connection one calculates
\begin{equation}
\int _{ \mathrm{TB} } \operatorname{Tr} (\Omega ^2 ) = 8 \pi ^2.
\end{equation} 
Therefore, by (\ref{ithm}),
\begin{equation}
\label{indexcount} 
\begin{split} 
\mathrm{index} (\slashed{D} _{ \mathcal{A} }) &
=\frac{ q (q + 1 )- p (p + 1 )}{2} .
\end{split} 
\end{equation} 
The absolute value of (\ref{indexcount})  is equal to the number of $L ^2 $ zero modes found in Section \ref{khjadsrewoui}, see equations (\ref{genzero}), (\ref{genzerod}). For a generic harmonic connection both $\T $ and $\T ^\dagger $ can  have non-trivial kernel, and the fact that (\ref{tbsoluni}), (\ref{ksolllljk})  give all the $L ^2 $ harmonic spinors rests on the explicit analysis of Section \ref{khjadsrewoui}. In the self-dual case  $\mathrm{Ker}(\T ^\dagger )=0 $ and (\ref{indexcount}) directly confirms that (\ref{sdansatz}) gives all the $L ^2 $ harmonic spinors.

\subsection{Taub-NUT}
The case of TN is discussed in \cite{Pope:1978ffa,Franchetti:2017hs}. We briefly recall the result.
The connection (\ref{tnconnect}) has the same asymptotic value it has on TB,  hence the fourth term in (\ref{ithm}) again evaluates to
$- \tfrac{1}{24} $. Using e.g.~the curvature of the spin connection one  calculates 
\begin{equation}
\int _ \mathrm{TN}  \operatorname{Tr} (\Omega ^2 ) = - 16 \pi ^2 .
\end{equation} 
By (\ref{ftnhjg}), (\ref{intftn}),
\begin{equation}
\int _ \mathrm{TN}  \mathcal{F} \wedge \mathcal{F} = 4 \pi ^2 \, \left( q + \frac{1}{2} \right)  ^2.
\end{equation} 
Since any harmonic form is self-dual,  $\mathrm{Ker} (\T ^\dagger )$ is trivial. Hence
\begin{equation}
\mathrm{index} (\slashed{D} _{ \mathcal{A} }) =  \mathrm{dim}  \left( \mathrm{Ker} (\T )\right)
=\frac{1}{2} q (q + 1 ),
\end{equation} 
in agreement with (\ref{tnhs}). Therefore, (\ref{tnsolhar}) gives all the $L ^2 $  harmonic spinors on TN.

\subsection{Euclidean Schwarzschild}
In this case the only non-vanishing term in (\ref{ithm}) is the second one \cite{Pope:1978ffa}.
For $\mathcal{F} $ given by (\ref{escurvv}), using (\ref{esvarintoiy}) we calculate
\begin{equation}
\label{indexcountes} 
\mathrm{index} (\slashed{D} _{ \mathcal{A} }) =\frac{1}{8 \pi ^2 }\int _{ \mathrm{ES} } \mathcal{F} \wedge \mathcal{F} 
= pq.
\end{equation} 
The absolute value of (\ref{indexcountes}) is equal  to the number of zero modes found in Section \ref{eszm}, see equations (\ref{esdim1}), (\ref{esdim2}). In the self-dual case $ q =p $  we can conclude to have found all the $L ^2 $ harmonic spinors, but otherwise we cannot exclude that an ansatz more general than (\ref{haresansatz}), (\ref{anses2}) could give other solutions.

\newpage 
\section{Conclusions}
\label{conclusions} 
It is time to discuss and interpret our results. As previously remarked, TB, TN and ES are Ricci-flat complete 4-manifolds with $SU (2)\times U (1) $ or $SO (3) \times U (1) $  as their isometry group. There are similarities in their asymptotic geometry (TN and TB), topology (ES and TB), structure of the  fixed points locus under the $U (1) $ action (ES and TB).

Among the three manifolds, TN has the simplest structure. The space of $L ^2 $ harmonic 2-forms is 1-dimensional and every harmonic form is self-dual. Connections with $L^2$ harmonic curvature which extend to the HHM compactification are classified by an integer $q$. Harmonic $L ^2 $ spinors belong to the direct sum of the   $SU (2) $ representation of dimension $n$ for $n$ ranging from $0$ to $|q |$. 
This simple structure can be related to the fact that TN is topologically trivial and equipped with a particularly rich geometrical structure --- it is a hyperk\"ahler manifold, a fact which by itself forces every $L ^2 $ harmonic form to be self-dual \cite{Hitchin:398393}. Topological considerations still arise, for example in the quantisation of $q$, due to the r\^ole played by the HHM compactification, given in this case by $\mathbb{C}  P ^2 $.

We note that the HHM compactification takes into account the geometry of the space. In fact for spaces of  ALF type such as those we have been considering, the circle fibres approach a finite length asymptotically, while the volume grows unboundedly in the other directions. The HHM compactification is obtained by shrinking to zero size the finite-size asymptotic circle fibres.

TB and ES both have the homotopy type of a 2-sphere, and their HHM compactification has 2-dimensional middle homology.  However, due to the differences in the topology of their fixed $r$ hypersurfaces, the  HHM compactifications are different,  the quadric $\mathbb{C}  P ^1 \times \mathbb{C}  P ^1 $ in the case of ES and $ \mathbb{C}  P ^2 \# \overline{ \mathbb{C}  P }^2 $ in the case of TB. The dimension of the space of $L ^2 $ harmonic forms is only sensitive to the dimension of the middle homology of the HHM compactification, and correspondingly both TB and ES have a 2-dimensional space of $L ^2 $ harmonic 2-forms, with  a 1-dimensional self-dual subspace. Connections extending to the HHM compactification with $L ^2 $ harmonic curvature are parametrised by two integers $p$, $q$. The more intricate topology of TB however manifests itself in the  relation between $\mathrm{d} \xi ^\flat $, $* \mathrm{d} \xi ^\flat  $ and  the Poincar\'e duals of a basis of 2-cycles, compare (\ref{fffftb}) with (\ref{ffffes}).

The differences between TB and ES  also emerge in the behaviour of harmonic spinors  under the $ SU (2) $  action. On TB $L ^2 $ harmonic spinors belong to the $SU (2) $ representation of dimension $n$, with $n$ varying in a range strictly related to the values of the integers $p$, $q$, see equations (\ref{unicondh}), (\ref{unicondk}). On ES instead,  harmonic spinors belong to a $SU (2) $ representation of fixed dimension $|p |$, with  degeneracy $|q |$. 

For all three spaces, the splitting of the space of harmonic $L ^2 $ spinors as a direct sum of $SU (2) $ representations can be related to the limiting values of the connection $\mathcal{A}$ as it approaches the $U (1) $ fixed point loci.
Let us take $\omega _0 =\frac{1}{2} \eta _3 $, the connection on the $U (1) $ bundle over $S ^2 $ with first Chern number one, as our reference value. In the case of  TN,   $\mathcal{A}$ vanishes on the nut $r =N $ and converges to $(q + \tfrac{1}{2} ) \omega _0  $ on the bolt $\Sigma _\infty $, see equation (\ref{tnconnect}), corresponding to the range (\ref{tnrangesu2})  of  allowed dimensionalities for the $SU (2) $ representations. In the case of TB, $\mathcal{A}$ ranges from 
$\left( p + \tfrac{1}{2}\right) \omega _0  $ at the bolt $r =2 N $ to $\left( q + \tfrac{1}{2}\right) \omega _0  $ at the bolt $\Sigma _\infty $, see equation (\ref{atb}), corresponding to the range (\ref{unicondh}) or (\ref{unicondk}) for the allowed dimensionalities of the $SU (2) $ representations. On ES, $\mathcal{A}$ has the same limit, $p \omega _0 $, see equation (\ref{esconn}), at both the bolt  $r =2 M $  and $\Sigma _\infty $, corresponding to the fact that harmonic spinors always belong to the $SU (2) $ representation of dimension $ |p |$.

In the cases of TB and TN  we have found all the $L ^2 $ harmonic spinors. The same can be said for ES if the Dirac operator is twisted by a self-dual connection, but for a generic $L ^2 $ harmonic connection the space of zero modes could be bigger. In all cases our results are consistent with the value of the index of $\slashed{D} _\mathcal{A} $. Interestingly, all the harmonic spinors that we have found are eigenstates of the chirality operator, that is either purely left- or right-handed.

\section*{Acknowledgements}
I would like to thank Tommaso Pacini and Bernd Schroers for interesting and insightful discussions, and the anonymous referee for useful comments which improved the presentation of this paper.

\newpage 
\appendix
\section{The twisted Dirac operator}
\label{twistdirac} 
We follow the general setup of \cite{Franchetti:2017hs} which we recall  here.
Let $\{ \gamma _\mu\} $, $ \mu =1, \ldots , 4 $, be Clifford generators,
\begin{equation}
\gamma _\mu \gamma _\nu + \gamma _\nu \gamma _\mu = - 2  \delta _{ \mu \nu } I _4 ,
\end{equation} 
where $I _4 $ is the $4 \times 4 $ identity matrix.
We take the generators  in the chiral form
\begin{equation}
\label{gammaschi} 
\gamma_a = \begin{pmatrix}
\mathbf{0}  & \ssigma _a  \\
- \ssigma _a  & \mathbf{0} 
\end{pmatrix} , \  a =1, 2, 3, \qquad 
\gamma _4 = \begin{pmatrix}
\mathbf{0}  & -i \mathbf{1}  \\
 -i  \mathbf{1}  & \mathbf{0} 
\end{pmatrix},
\end{equation} 
where $\mathbf{1}, \mathbf{0}$ are the  $2 \times 2 $ identity and null matrices and $\{\ssigma _a\} $ the Pauli matrices.
Dirac spinors which are eigenvectors of the chirality operator $- \gamma _1 \gamma _2 \gamma _3 \gamma _4 $ with eigenvalue $+1$ (respectively $-1$) are called left-handed (right-handed). In the chiral representation (\ref{gammaschi}), the third and fourth (first and second) components of a left-handed (right-handed) Dirac spinor vanish.

The twisted Dirac operator $\slashed{D} _\mathcal{A} $ associated to an orthonormal coframe $\{e_\mu\} $, its dual frame $\{ E _\mu \} $  and the Abelian real-valued connection $\mathcal{A} $ is, see e.g.~\cite{Lawson:276634},
\begin{equation}
\label{sled} 
\slashed{D}  _\mathcal{A}  
= \gamma _\mu  \left[ \big(E _\mu  +i \mathcal{A}  (E _\mu )\big) I _4  - \frac{1}{8} [ \gamma  _\rho , \gamma _\sigma ] \, \omega _{ \rho \sigma } (E _\mu ) \right].
\end{equation} 
The non-twisted Dirac operator $\slashed{D} $  is obtained by setting  $\mathcal{A} = 0 $.

In terms of the left-invariant 1-forms on $SU (2) $,
\begin{equation}
\begin{split}
\eta _1 &= \sin \psi  \, \mathrm{d} \theta - \cos \psi \sin \theta \, \mathrm{d} \phi,\\
\eta _2 &= \cos \psi \, \mathrm{d} \theta + \sin \psi \sin \theta \, \mathrm{d} \phi ,\\
\eta _3 &= \mathrm{d} \psi + \cos \theta \, \mathrm{d} \phi,
\end{split}
\end{equation} 
with $\theta \in[0, \pi ] $, $\phi \in[0, 2 \pi )$, $\psi \in[0, 4 \pi )$,
a  Bianchi IX metric has the form
\begin{equation}
g _{ IX }= f ^2 \mathrm{d} r ^2 + a ^2  \eta _1 ^2 + b ^2  \eta ^2 _2 + c ^2  \eta _3 ^2
\end{equation}
with $a$, $b$, $c$, $f$ functions of the transverse coordinate $r$ only.  If $a =b $, so that the metric acquires a $U (1) $ isometry generated by the Killing vector field $\partial / \partial \psi $, the metric is said to be bi-axial. In the following we write $\dot{\ } $ for $f ^{-1} \mathrm{d} / \mathrm{d} r  $, and $^\prime $ for $\mathrm{d} / \mathrm{d} r $. 

We take the orthonormal coframe
\begin{equation}
e ^1 =a \eta _1 , \quad e ^2 =b \eta _2 , \quad e ^3 =c \eta _3 , \quad e ^4 = - f \mathrm{d}r .
\end{equation} 
In our conventions Latin indices vary in the range $\{1,2,3\} $ and Greek indices vary in the range $\{1,2,3,4\} $. The Einstein summation convention is enforced but,  since we are working with an orthonormal coframe, we do not distinguish upper indices from lower ones.
We denote the orthonormal frame dual to $\{ e ^\mu \} $ by $\{ E _\mu \} $. Note that
\begin{equation}
E _1 =a ^{-1}  X _1 , \quad E _2 =b ^{-1}  X _2 , \quad E _3 = c^{-1}  X _3, \quad E _4 =- f ^{-1} \partial _r ,
\end{equation} 
where
\begin{equation}
\begin{split}
X _1 &= \sin \psi\,  \partial _\theta + \frac{\cos \psi }{\sin \theta } \left( \cos \theta \, \partial _\psi - \partial _\phi \right) ,\\
X _2 &= \cos \psi\,  \partial _\theta - \frac{\sin  \psi }{\sin \theta } \left( \cos \theta \, \partial _\psi - \partial _\phi \right) ,\\
X _3 &= \partial _\psi
\end{split}
\end{equation} 
are the left-invariant vector fields on $SU (2) $ dual to the forms $\{ \eta _i \} $.

Define
\begin{equation}
\label{ABC} 
A  = \frac{ b ^2 + c ^2 - a ^2  } {2bc }, \quad 
B  =\frac{c ^2 + a ^2 - b ^2 }{2ca}, \quad 
C   =\frac{a ^2 + b ^2 - c ^2 }{2ab}.
\end{equation} 
and 
\begin{equation}
D _i = X _i + i \mathcal{A} (X   _i),\  i =1 ,2,3, \quad D _{ \pm } = D _1 \pm i D _2 .
\end{equation} 
It can be shown in \cite{Jante:2014ho,Franchetti:2017hs} that for a Bianchi IX metric $\slashed{D}  _\mathcal{A}   $  can be written
\begin{equation}
\slashed{ D } _\mathcal{A} =
 \begin{pmatrix}
 \mathbf{0}  & \T ^\dagger    \\
\T   & \mathbf{0} 
\end{pmatrix}
\end{equation} 
with
\begin{align}
\label{tt} 
\T   & 
=\left[ \frac{i\partial _r}{f} + \mathcal{A} (E _4  )  + \frac{i}{2} \left( \frac{\dot a  }{a} + \frac{\dot b  }{b} + \frac{\dot c }{c}\right)  \right] \mathbf{1} 
+ i  \, \dbb
\\ \nonumber & = \left[ \frac{i\partial _r}{f} + \mathcal{A} (E _4 )  + \frac{i}{2} \left( \frac{\dot a + A }{a} + \frac{\dot b + B }{b} + \frac{\dot c + C }{c}\right)  \right] \mathbf{1} 
- \frac{ \ssigma _1 D _1}{a} 
- \frac{ \ssigma _2 D _2}{b} - \frac{ \ssigma _3  D_ 3 }{c} ,\\
\label{ttd} 
\T  ^\dagger &
=\left[ \frac{i\partial _r}{f}  + \mathcal{A} (E _4  )   + \frac{i}{2} \left( \frac{\dot a  }{a} + \frac{\dot b  }{b} + \frac{\dot c }{c}\right)  \right] \mathbf{1} 
-i \,   \dbb\\ \nonumber &
= \left[ \frac{i\partial _r}{f}  + \mathcal{A} (E _4  )   + \frac{i}{2} \left( \frac{\dot a - A }{a} + \frac{\dot b - B }{b} + \frac{\dot c - C }{c}\right) \right] \mathbf{1}   + \frac{ \ssigma _1 D _1}{a} 
+ \frac{ \ssigma _2 D _2}{b} + \frac{ \ssigma _3 D _3 }{c},
\end{align} 
where
\begin{equation}
\dbb = 
 i \left(  \frac{ \ssigma _1 D _1 }{a} + \frac{ \ssigma _2 D _2 }{b} + \frac{ \ssigma _3 D _3 }{c}  \right)  
 +  \frac{\mathbf{1} }{2} \left( \frac{A }{a} + \frac{B }{b} + \frac{C }{c}  \right).
\end{equation} 
The operator $\T ^\dagger   $ is the  formal adjoint of  $\T  $.

In the case of a bi-axial Bianchi IX metric $( a =b \Rightarrow A =B $), setting
\begin{equation}
\lambda =c/a,
\end{equation} 
we calculate
\begin{equation}
\begin{split} 
&\frac{1 }{2} \left( \frac{A }{a} + \frac{B }{b} + \frac{C }{c}  \right)=\frac{1}{2a} \left( \frac{2 + \lambda ^2 }{2 \lambda } \right) ,\\
& i \left(  \frac{ \ssigma _1 D _1 }{a} + \frac{ \ssigma _2 D _2 }{b} + \frac{ \ssigma _3 D _3 }{c}  \right) 
=\begin{pmatrix}
 i D _3 /c & i D _-/a \\
 i D _+ /a & -i D _3 /c
\end{pmatrix} =
\frac{1}{2a}\begin{pmatrix}
2i D _3 / \lambda  & 2i D _-  \\
2i D _+  & - 2i D _3 / \lambda 
\end{pmatrix} ,
\end{split} 
\end{equation} 
so that
\begin{equation}
\dbb  =\frac{1}{2a}\begin{pmatrix}
2i D _3 / \lambda  & 2i D _-  \\
2i D _+  & - 2i D _3 / \lambda 
\end{pmatrix} + \frac{\mathbf{1} }{2a} \left( \frac{2 + \lambda ^2 }{2 \lambda } \right) 
\end{equation} 
and
\begin{align} 
\T&=  \frac{i}{f} \left( \partial _r + \frac{a ^\prime }{a} + \frac{c ^\prime }{2c} + \mathcal{A} (E _4  ) \right) \mathbf{1}  + i \dbb,\\
\T ^\dagger  &
= \frac{i}{f} \left( \partial _r  +   \frac{ a ^\prime  }{a} + \frac{ c ^\prime  }{2c} + \mathcal{A} (E _4  )\right) \mathbf{1} - i \dbb.
\end{align} 

\newpage 
\section{Eigenvectors of $\dbbb $ leading to no $L ^2 $ solutions}
\label{othereigen} 
In this appendix we show that eigenvectors of the form (\ref{genegv}) lead to no non-trivial $L ^2 $ harmonic spinors on TB. Our starting point is the solution (\ref{sol1aa}) of the Dirac equation.

Substituting the eigenvalue (\ref{genegv}) of  $\dbbb $ for $\Lambda$  in  (\ref{sol1aa}),  with $a$, $b$, $c$, $f$ given by   (\ref{tbvalues}), $\p $ by (\ref{ptiltb}) and $A$, $B$ satisfying (\ref{abval}), we obtain
\begin{equation}
\label{asdkljerwiuy} 
\begin{split} 
&h= \frac{ C}{ \sqrt{ r + N } }\, \left( r - 2 N \right) ^{ -1/4 }\left( r - N /2 \right) ^{ -1/4 } 
 \exp \left[ \pm \frac{1}{2 N } 
 \int  \frac{ \sqrt{ \gamma} \, \mathrm{d} r}{(r- 2 N ) (2r- N )  } \right] ,
\end{split} 
\end{equation} 
with $C$ an arbitrary constant.
The sign choice corresponds to the one in (\ref{genegv}), and $\gamma$ is 
\begin{equation}
\begin{split} 
\gamma &=16 N  ^2 (j-m)(j+m+1) (r ^2 - 5Nr/2 + N ^2  ) + \\ &
+  \frac{1}{4} \Big(N^2 (4 m+2 q+3)+r^2 (-4 m+2
   q-1)+N r (3 p-5 q-1)\Big)^2 .
   \end{split} 
\end{equation} 
The integral in (\ref{asdkljerwiuy})  can be expressed in terms  of the roots of a polynomial of 4th order in $r$ whose coefficients depend on the parameters $m $, $j$, $p$, $q$. However, that is not very useful. Instead, to determine square integrability at large $r$ and near the bolt it is enough to estimate the behaviour for $r \gg 1 $ and  $ r \simeq 2 N $.

For large $r$ we obtain
\begin{equation}
\begin{split}
&\exp  \left[ \pm \frac{1}{2 N } 
 \int  \frac{ \sqrt{ \gamma} \, \mathrm{d} r}{(r- 2 N ) (2r- N )  } \right] 
 \simeq  \exp\left[\pm \left( \frac{ |1 + 4m - 2q | }{8N} \right)  r\right]   ,
\end{split} 
\end{equation} 
while for $r \simeq 2 N $ we obtain
\begin{equation}
\exp  \left[ \pm \frac{1}{2 N } 
 \int  \frac{ \sqrt{ \gamma} \, \mathrm{d} r}{(r- 2 N ) (2r- N )  } \right] 
 \simeq   \left( r- 2 N \right) ^{\pm \frac{1}{4}  |1 + 4m-2p |  } .
\end{equation} 
Since $p, q, 2m \in \mathbb{Z}  $, the expressions $1 + 4m - 2q $, $1 + 4m-2p $  never vanish. Square integrability  at infinity and near the bolt require incompatible sign choices, thus we conclude that the eigenvector choice (\ref{genegv}) leads to no $L ^2 $ harmonic spinors. For the solution (\ref{sol2aa}) similar computations lead to the same conclusion.

\newpage 
\printbibliography
\end{document}